\documentstyle[epsfig,graphicx]{cupconf}

\newcommand{\be}{\begin{equation}}
\newcommand{\ee}{\end{equation}}

\def\mHH{H_2}
\def\mH2p{H_2^+}
\def\mHp{H^+}
\def\mHm{H^-}

\newcommand{\Lya}{Ly$\alpha\ $}
\def\lya{Ly$\alpha\ $}

\newcommand{\HH}{H$_2$}

\newcommand{\etal}{et~al.\ }
\def\kms{\,{\rm km\,s^{-1}}}
\def\kmsmpc{\,{\rm km\,s^{-1}\,Mpc^{-1}}}
\def\msun{\,{\rm M_\odot}}

\def\spose#1{\hbox to 0pt{#1\hss}}
\def\lta{\mathrel{\spose{\lower 3pt\hbox{$\mathchar"218$}} \raise 2.0pt\hbox{$\mathchar"13C$}}}
\def\gta{\mathrel{\spose{\lower 3pt\hbox{$\mathchar"218$}} \raise 2.0pt\hbox{$\mathchar"13E$}}}

\def\HI{\hbox{H~$\scriptstyle\rm I$}}
\def\HII{\hbox{H~$\scriptstyle\rm II$}}
\def\HeI{\hbox{He~$\scriptstyle\rm I$}}
\def\HeII{\hbox{He~$\scriptstyle\rm II$}}
\def\HeIII{\hbox{He~$\scriptstyle\rm III$}}

\def\nHI{{\rm HI}}
\def\nH{{\rm H}}

\def\vir{{\rm vir}}

\newcommand{\rvir}{r_{\rm vir}}

\newcommand{\msub}{M_{\rm sub}}
\newcommand{\Tvir}{T_{\rm vir}}

\newcommand{\rtwo}{r_{200}}
\newcommand{\mtwo}{M_{200}}
\newcommand{\tg}{T_e}
\newcommand{\tr}{T}

\def\s{\sigma}
\newcommand\sigc{\s_{\log c}}
\def\medc{{c}_*}

\title[Early Galaxy Formation]{The Astrophysics of Early Galaxy Formation}

\author[Piero Madau]{Piero Madau\\ Department of Astronomy and Astrophysics\\ University 
of California Santa Cruz\\
}

\begin{document}

\maketitle

\section{Preamble}

\medskip
\noindent Hydrogen in the universe recombined about half a million years after the 
Big Bang, and cooled down to a temperature of a few kelvins until the
first non-linearities developed, and evolved into stars, galaxies,
and black holes that lit up the universe again. In currently popular cold 
dark matter flat cosmologies ($\Lambda$CDM), some time beyond a redshift of
10 the gas within halos with virial temperatures $T_\vir\gta 10^4\,$K -- or,
equivalently, with masses $M\gta 10^8\, [(1+z)/10]^{-3/2}\,\msun$ -- cooled rapidly due
to the excitation of
hydrogen \Lya and fragmented. Massive stars formed with some initial mass
function (IMF), synthesized heavy elements, and exploded as
Type II supernovae after a few $\times 10^7\,$yr, enriching the
surrounding medium: these subgalactic stellar systems, aided perhaps by an
early population of accreting black holes in their nuclei, generated the ultraviolet
radiation and mechanical energy that contributed to the
reheating and reionization of the cosmos. It is widely believed that collisional excitation
of molecular hydrogen may have allowed gas in even smaller systems -- virial
temperatures of a thousand K, corresponding to masses around
$5\times 10^5\,[(1+z)/10]^{-3/2}\,\msun$ --  to cool and form stars at even earlier times
(Couchman \& Rees 1986; Haiman \etal 1996; Tegmark \etal 1997; Abel \etal 2000, 2002;
Fuller \& Couchman 2000; Bromm \etal 2002; Reed \etal 2005; Kuhlen \& Madau 2005).
Throughout the epoch of structure
formation, the all-pervading intergalactic medium (IGM), which contains most of the ordinary
baryonic material left over from the Big Bang, becomes clumpy under the
influence of gravity, and acts as a source for the gas that gets
accreted, cools, and forms stars within subgalactic fragments, and as a sink
for the metal enriched material, energy, and radiation which they eject.
The well-established existence of heavy elements like carbon and silicon in the \lya
forest clouds at $z=2-6$ (Songaila 2001; Pettini \etal 2003; Ryan-Weber \etal 2006)
may be indirect evidence for such an early episode of pregalactic star formation.
The recently released {\it Wilkinson Microwave Anisotropy Probe (WMAP)} 3-year data 
require the universe
to be fully reionized by redshift $z_r=11\pm 2.5$ (Spergel \etal 2006), another indication
that significant star-formation activity started at very early cosmic times.

The last decade has witnessed great advances in our understanding of the
high redshift universe, thanks to breakthroughs achieved
with satellites, 8-10m class telescopes, and cosmic microwave background 
(CMB) experiments. Large surveys such as the {\it Sloan
Digital Sky Survey} (SDSS), together with the use of novel instruments and
observational
techniques have led to the discovery of galaxies and quasars at redshifts
in excess of 6. At the time of writing, 9 quasars have already been found
with $z>6$ (Fan 2006), and one actively star-forming has been spectroscopically 
confirmed at $z=6.96$ (Iye \etal 2006). These sources probe an epoch 
when the universe was $<7\%$ of its current
age. {\it Keck} and {\it VLT} observations of redshifted \HI\ \Lya (`forest')
absorption have been shown to be sensitive probe of the distribution of
gaseous matter in the universe (see Rauch 1998 for a review). Gamma-ray bursts have recently displayed
their potential to replace quasars as the preferred probe
of early star formation and chemical enrichment: GRB050904, the most distant
event known to date, is at $z=6.39$ (Haislip \etal 2006).
The underlying goal of
all these efforts is to understand the growth of cosmic structures,
the properties of galaxies and their evolution, and ultimately to map
the transition from the cosmic ``dark age'' to a ionized universe
populated with luminous sources.

Progress has been equally significant on the theoretical side. The key idea
of currently popular
cosmological scenarios, that primordial density fluctuations grow by
gravitational instability driven by collisionless CDM, 
has been elaborated upon and explored in detail through
large-scale numerical simulations on supercomputers, leading to a
hierarchical (`bottom-up') scenario of structure formation. In this model,
the first objects to form are on subgalactic scales, and merge to
make progressively bigger structures (`hierarchical clustering').
Ordinary matter in the universe follows the dynamics dictated by the
dark matter until radiative, hydrodynamic, and star formation processes
take over. According to these
calculations, a truely inter- and proto-galactic medium (the main
repository of baryons at high redshift) collapses under the influence
of dark matter gravity into flattened and filamentary structures, which
are seen in absorption against background QSOs.
Gas condensation in the first
baryonic objects is possible through the formation of H$_2$ molecules,
which cool via roto-vibrational transitions down to temperatures of a few
hundred kelvins. In the absence of a UV
photodissociating flux and of ionizing X-ray radiation, three-dimensional
simulations of early structure formation show that the fraction of
cold, dense gas available for accretion onto seed black holes or star formation
exceeds 20\% for halos more massive than $10^6\,\msun$ already at redshifts 20
(Machacek \etal 2003; Yoshida \etal 2003).

In spite of some significant achievements in our understanding of the
formation of cosmic structures, there are still many challenges facing
hierarchical clustering theories. While quite successful in matching the
observed large-scale density distribution (like, e.g., the properties
of galaxy clusters, galaxy clustering, and the statistics of the \Lya
forest), CDM simulations appear to
produce halos that are too centrally concentrated compared to the mass
distribution inferred from the rotation curves of (dark matter-dominated)
dwarf galaxies, and to predict too many dark matter subhalos compared to
the number of dwarf satellites observed within the Local
Group (Moore \etal 1999; Klypin \etal 1999). Another perceived difficulty (arguably
connected with the ``missing satellites problem'', e.g. Bullock \etal 2000) is our inability
to predict when and how the universe was reheated and reionized, i.e. to
understand the initial conditions of the galaxy formation process and the basic
building blocks of today's massive baryonic structures. We know that at least some galaxies
and quasars had already formed when the universe was less than $10^9\,$ yr
old. But when did the first luminous clumps form, was star formation
efficient in baryonic objects below the atomic cooling mass, and what was the
impact of these early systems on the surrounding intergalactic gas?
The crucial processes of star formation and ``feedback'' (e.g. the effect of the energy input
from the earliest generations of sources on later ones)
in the nuclei of galaxies are still poorly understood.
Accreting black holes can release large amounts of energy to their
surroundings, and may play a role in regulating the thermodynamics of
the interstellar, intracluster, and intergalactic medium. The detailed
astrophysics of these processes is, however, unclear.
Although we may have a sketchy history of
the production of the chemical elements in the universe, we know
little about how and where exactly they were produced and how they are
distributed in the IGM and in the intracluster medium.
Finally, where are the first stars and their remnants now, and why are the
hundreds of massive satellites predicted to survive today in the Milky Way halo dark?

In these lectures I will describe some of the basic tools that have been developed to study 
the dawn of galaxies, the lessons learned, and summarize our current understanding of 
the birth of the earliest astrophysical objects.

\section{The Dark Ages}
 
\subsection{Cosmological preliminaries}

Recent CMB experiments, in conjunction with new measurements of the large-scale structure
in the present-day universe, the SNIa Hubble diagram, and other observations, have lead
to a substantial reduction in the uncertanties in the parameters describing the 
$\Lambda$CDM concordance cosmology. We appear to be living in a flat ($\Omega_m+
\Omega_\Lambda=1$) universe dominated by a 
cosmological constant and seeded with an approximately scale-invariant primordial spectrum
of Gaussian density fluctuations. A $\Lambda$CDM cosmology with $\Omega_m=0.24$, 
$\Omega_\Lambda=0.76$, 
$\Omega_b=0.042$, $h=H_0/100\,\kmsmpc=0.73$, $n=0.95$, and $\sigma_8=0.75$ is 
consistent with the best-fit parameters from the {\it WMAP} 3-year data 
release (Spergel et al. 2006). Here $\Omega_m=\rho^0_m/\rho_c^0$\footnote{Hereafter
densities measured at the present epoch will be indicated either by the superscript
or the subscript `0'.}\ is the present-day 
matter density (including cold dark matter as well as a contribution $\Omega_b$ from baryons) 
relative to the critical density $\rho_c^0=3 H_0^2/8 \pi G$, $\Omega_\Lambda$ is the
vacuum energy contribution, $H_0$ is the Hubble constant today, $n$ is 
the spectral index of the matter power spectrum at inflation, and $\sigma_8$ normalizes 
the power spectrum: it is the root-mean-square amplitude of mass fluctuations in 
a sphere of radius $8 h^{-1}\,$Mpc. The lower values of $\sigma_8$ 
and $n$ compared to {\it WMAP} 1-year  results (Spergel et al. 2003) have the effect of delaying 
structure formation and reducing small-scale power. In this cosmology, and at the 
epochs of interest here, the expansion rate $H$ evolves according to the Friedmann equation
\be
H={1\over a}{da\over dt}=H_0 [\Omega_m (1+z)^3+\Omega_\Lambda]^{1/2}, 
\ee
where $z$ is the redshift. Light emitted by a source at time $t$ is observed today 
(at time $t_0$) with a redshift $z\equiv 1/a(t)-1$, where $a$ is the cosmic scale
factor [$a(t_0)\equiv 1$]. The age of the (flat) universe today is   
\be  
t_0 = \int_0^\infty \frac{dz'}{(1+z') \, H(z')}=
{2\over 3H_0\sqrt{\Omega_\Lambda}}\ln\left[{\sqrt{\Omega_\Lambda}+1
\over \sqrt{\Omega_m}}\right]=1.025\,H_0^{-1}=13.7~{\rm Gyr}
\ee 
(Ryden 2003). At high redshift, the universe approaches the 
Einstein-de Sitter behaviour, $a\propto t^{2/3}$, and its age is given by  
\be 
t(z) \approx {2\over 3H_0\sqrt{\Omega_m}}\,(1+z)^{-3/2}.
\ee
The average baryon density today is $n_b^0=2.5 \times 10^{-7}$ cm$^{-3}$, and the 
hydrogen density is $n_{\rm H}=(1-Y_p)\,n_b=0.75\,n_b$, where $Y_p$ is the primordial 
mass fraction of helium. The photon density is $n_\gamma^0=(2.4/\pi^2)\,
(k_BT_0/\hbar c)^3=400\,{\rm cm^{-3}}$, where $T_0=2.728\pm 0.004$ K
(Fixsen et al. 1996), and the cosmic baryon-to-photon ratio is then
$\eta=n_b/n_\gamma=6.5\times 10^{-10}$.
The electron scattering optical depth to redshift $z_r$ is given by
\be 
\tau_e(z_r)=\int_0^{z_r} dz\,n_e\sigma_T c{dt\over dz}=
\int_0^{z_r} dz\,\frac{n_e(z)\sigma_T c}{(1+z)\,H(z)}
\ee
where $n_e$ is the electron density at redshift $z$ and $\sigma_T$ the Thomson
cross-section. Assuming a constant electron fraction $x_e\equiv n_e/n_{\rm H}$
with redshift, and neglecting the vacuum energy contribution in $H(z)$, this 
can be rewritten as
\be 
\tau_e(z_r)=\frac{x_e n_{\rm H}^0 \sigma _Tc}{H_0\sqrt{\Omega_m}}\int_0^{z_r} dz
\sqrt{1+z}=0.0022\,[(1+z_r)^{3/2}-1].
\ee 
Ignoring helium, the 3-year {\it WMAP} polarization data requiring $\tau_e=0.09\pm0.03$
are consistent with a universe in which $x_e$ changes from essentially
close to zero to unity at $z_r=11\pm 2.5$, and $x_e\approx 1$ thereafter. 

Given a population of objects with proper number density $n(z)$ and geometric cross-section 
sigma $\Sigma(z)$, the incremental probablity $dP$ that a line of sight will intersect
one of the objects in the redshift interval $dz$ at redshift $z$ is
\be 
dP=n(z)\Sigma(z)c{dt\over dz}dz=n(z)\Sigma(z)c{dz\over (1+z)H(z)}.
\ee 

\subsection{Physics of recombination}

Recombination marks the end of the plasma era, and the beginning of the era of
neutral matter. Atomic hydrogen has an ionization potential of $I=13.6$ eV: it takes 
10.2 eV to raise an electron from the ground state to the
first excited state, from which a further 3.4 eV will free it,
\be
\gamma + {\rm H}\longleftrightarrow p+e^-.
\ee
According to Boltzmann statistics, the number density distribution of non-relativistic 
particles of mass $m_i$ in thermal equilbrium is given by
\be 
n_i \,=\, g_i \left({m_i k_BT \over 2\pi \hbar^2}\right)^{3/2} \,
{\rm exp}\left({\mu_i-m_ic^2 \over k_BT}\right),
\label{eqboltzmann}
\ee 
where $g_i$ and $\mu_i$ are the statistical weight and chemical potential of the species.
When photoionization equibrium also holds, then $\mu_{e}+\mu_p=\mu_{\rm H}$. 
Recalling that $g_e=g_p=0.5$ and $g_{\rm H}=2$, one then obtains the Saha equation:
\be 
{n_e n_p \over n_\nH n_\nHI}
={x_e^2 \over 1-x_e}={(2\pi m_e k_B T)^{3/2} \over n_\nH(2\pi\hbar)^3}
\, {\rm exp}\left(-{I \over k_B T}\right),
\label{eqsaha}
\ee 
which can be rewritten as
\be 
\ln\left({x_e^2\over 1-x_e}\right)=52.4-1.5\,\ln(1+z)-58,000\,(1+z)^{-1}.
\ee 
The ionization fraction goes from 0.91 to 0.005
as the redshift decreases from 1500 to 1100, and the temperature
drops from 4100 K to 3000 K. The elapsed time is less than 200,000 yr.      
 
Although the Saha equation describes reasonably well the initial
phases of the departure from complete ionization, the assumption of
equilibrium rapidly ceases to be valid in an expanding universe, and 
recombination freezes out. The residual electron fraction can be estimated 
as follows. The rate at which electrons recombine with protons is
\be 
{dn_e\over dt}=-\alpha_B n_e n_p\equiv -{n_p\over t_{\rm rec}},
\ee 
where $t_{\rm rec}$ is the characteristic time for recombination and
$\alpha_B$ is the radiative recombination coefficient. This is the product of
the electron capture cross-section $\sigma_n$ and the electron velocity
$v_e$, averaged over a thermal distribution and summed over all excited
states $n\ge 2$ of the hydrogen atom, $\alpha_B=\sum \langle \sigma_n v_e\rangle$.
The radiative recombination coefficient is well approximated  by the fitting formula
\be
\alpha_B=6.8\times 10^{-13}\,T_3^{-0.8}~~{\rm
cm^3~s^{-1}}=1.85\times 10^{-10}\,(1+z)^{-0.8}~~{\rm cm^3~s^{-1}},
\ee 
where $T_3\equiv T/3000$ K and in the second equality I used $T=T_0(1+z)$.
When the recombination rate falls below the expansion rate, i.e. when
$t_{\rm rec}>1/H$, the formation of neutral atoms ceases and the remaining
electrons and protons have negligible probability for combining with each other:
\be 
t_{\rm rec}H={H_0\sqrt{\Omega_m}\over x_en_\nH^0\alpha_B}\,(1+z)^{-3/2}={0.0335\over x_e}
(1+z)^{-0.7}={2.5\times 10^{-4}\over x_e},
\ee 
where I assumed $z=1100$ in the third equality.
\begin{figure}
\begin{center}
\includegraphics[width=.49\textwidth]{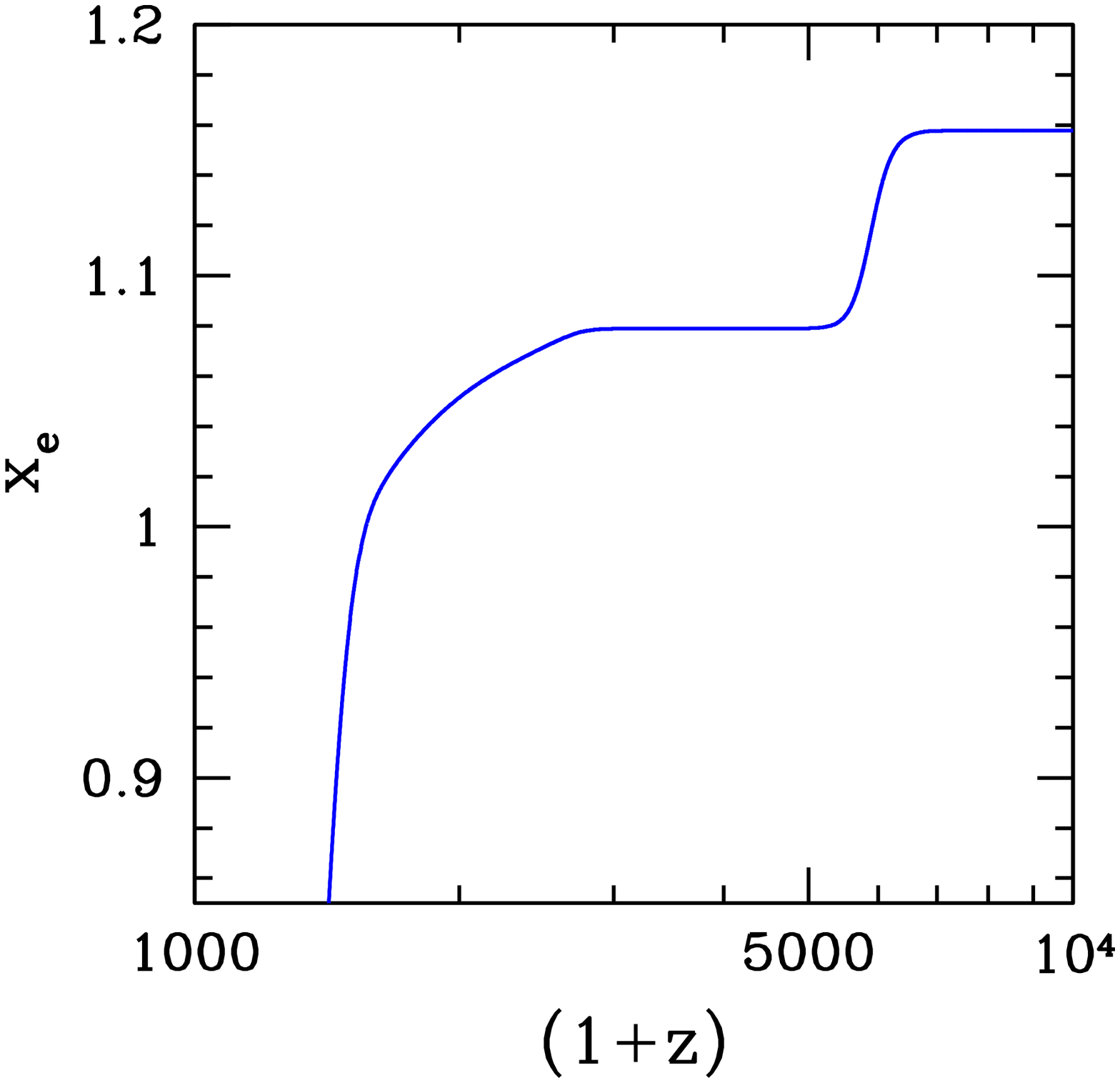}
\includegraphics[width=.48\textwidth]{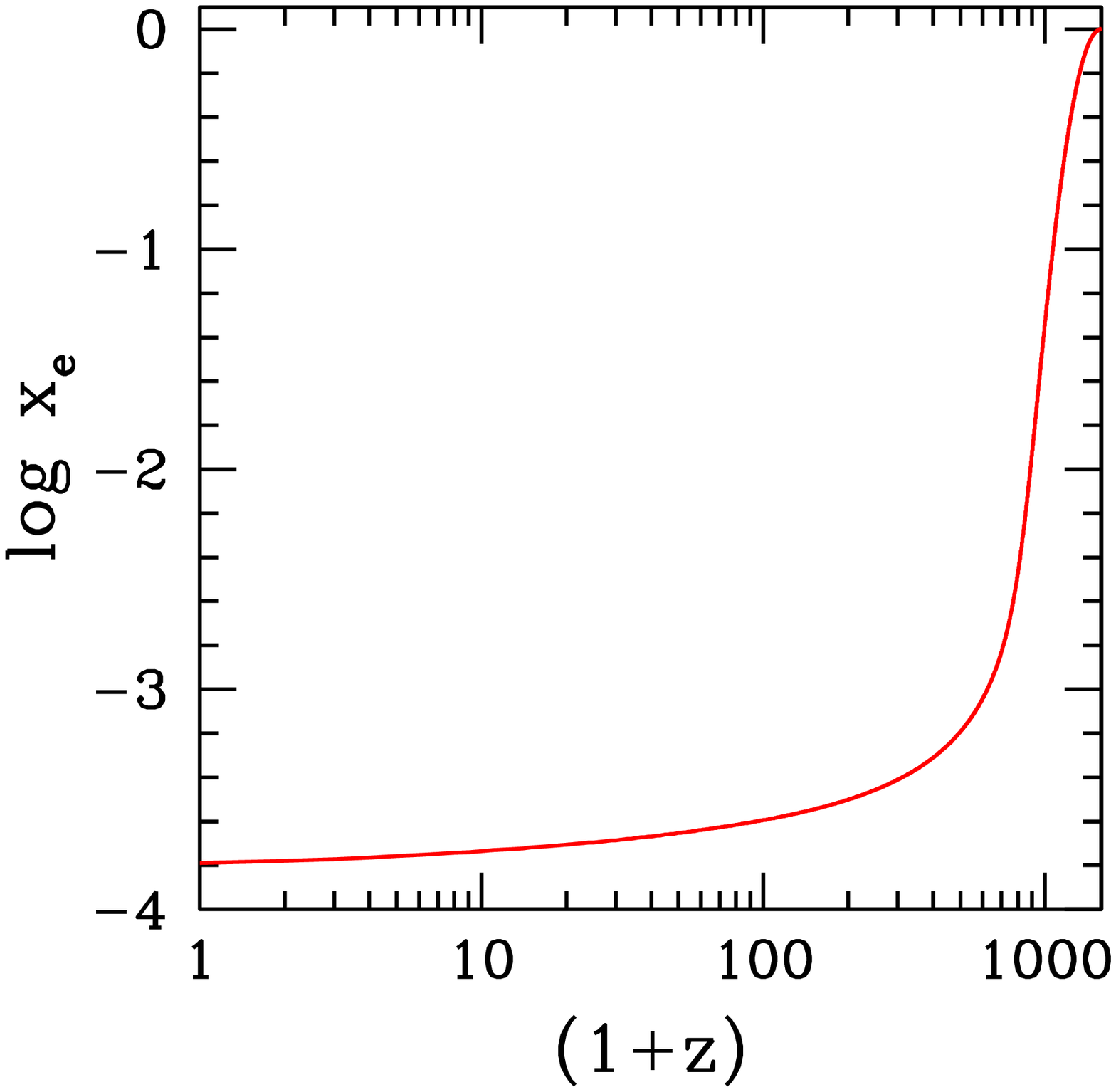}
\end{center}
\caption[]{Helium and hydrogen recombination for a {\it WMAP}
3-year cosmology. The step at earlier
times in the left panel is due to the recombination of \HeIII\ into
\HeII. We used the code RECFAST (Seager \etal 1999) to compute the
electron fraction $x_e$. Note that the residual electron 
fraction (determined by the condition $t_{\rm rec}H=1$) scales with cosmological
parameters as $x_e\propto \sqrt{\Omega_m}/\Omega_b h$.
\label{fig:recombination}
}
\end{figure}
The recombination of hydrogen in an expanding universe is actually delayed  by 
a number of subtle effects that are not taken into account  in the above formulation. 
An $e^-$ captured to the ground state of atomic hydrogen produces a photon 
that immediately ionizes another atom,
leaving no net change. When an $e^-$ is captured instead to an excited state,
the allowed decay to the ground state produces a resonant Lyman series photon.
These photons have large capture cross-sections and put atoms in a high energy 
state that is easily photoionized again, thereby annulling the effect. That leaves two 
main routes to the production of atomic hydrogen: 1) two-photon decay from the $2s$ level to 
the ground state; and (2) loss of Ly$\alpha$ resonance photons by the 
cosmological redshift. The resulting recombination history was derived by 
Peebles (1968) and Zel'dovich et al. (1969). 

In the redshift range $800<z<1200$, the fractional ionization varies rapidly and 
is given approximately by the fitting formula
\be 
x_e=0.042\,\left({z\over 1000}\right)^{11.25}.
\ee 
This is a fit to a numerical output from the code RECFAST -- an
improved calculation of the recombination of \HI, \HeI, and \HeII\ in the early universe 
involving a line-by-line treatment of each atomic level (Seager \etal 1999).
Using this expression, we can again compute the optical depth of the universe 
for Thomson scattering by free electrons:
\be 
\tau_e(z)=\int_0^z dz\,\frac{x_e n_\nH\sigma_T c}{(1+z)\,H(z)}
\approx 0.3\,\left({z\over 1000}\right)^{12.75}.
\ee 
This optical depth is unity at $z_{\rm dec}=1100$. From the optical depth we can compute the 
{\it visibility function} $P(\tau)$, the probability that a photon was last 
scattered in the interval $(z,z+dz)$. This is given by 
\be 
P(\tau)=e^{-\tau}\,{d\tau\over dz}=0.0038\,\left({z\over 1000}\right)^{11.75}\,
e^{-0.3(z/1000)^{12.75}},
\ee
and has a sharp maximum at $z=z_{\rm dec}$ and a width of $\Delta z\approx 100$.
The finite thickness of the last scattering surface has important observational 
consequences for CMB temperature fluctuations on small scales. 
\begin{figure}
\vspace{0.cm}
\centerline{\psfig{figure=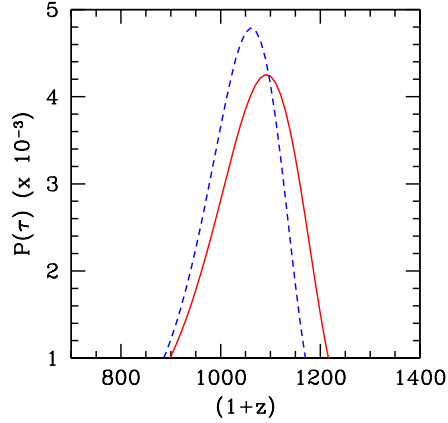,width=0.49\textwidth}}
\caption{Visibility function. {\it Solid line:} RECFAST with {\it WMAP} 3-year cosmological
parameters. {\it Dashed line:} Jones \& Wyse (1985). 
}
\vspace{0.5cm}
\end{figure}

\subsection{Coupling of gas and radiation}

The CMB plays a key role in the early evolution of 
structures. First, it sets the epoch of decoupling, when baryonic matter becomes free to 
move through the radiation field to form the first generation of gravitationally 
bound systems. Second, it fixes the matter temperature that in turns 
determines the Jeans scale of the minimum size of the first bound objects.

Following Peebles (1993), consider an electron $e^-$ moving at non-relativistic speed
$v\ll c$ through the CMB. In the $e^-$ rest-frame, the CMB temperature
measured at angle $\theta$ from the direction of motion is
\be 
T(\theta)=T\left(1+{v\over c}\cos\theta\right).
\ee 
The radiation  energy density per unit volume per unit solid angle $d\Omega=
d\phi d\cos\theta$ around $\theta$ is
\be 
du=[a_BT^4(\theta)]~{d\Omega\over 4\pi} 
\ee 
(here $a_B$ is the radiation constant), and the net drag force (i.e. the component of the 
momentum transfer along the direction of motion, integrated over all 
directions of the radiation) felt by the electron is
$$
F=\int_{4\pi}\sigma_Tdu\cos\theta=\int
\sigma_T (a_BT^4) \left(1+{v\over c}\cos\theta\right)^4 \cos\theta {d\Omega\over 4\pi}
$$ 
\be 
~~~~~~~~~~~={4\over 3} {\sigma_T a_BT^4\over c}v.  
\ee 
This force
will be communicated to the protons through electrostatic coupling.
The formation of the first gravitationally bound systems
is limited by radiation drag, as the drag force per baryon is $x_eF$, where $x_e$ is
the fractional
ionization. The mean force divided by the mass $m_p$ of a hydrogen atom gives  
the deceleration time of the streaming motion:
\be
t_s^{-1}={1\over v} {dv\over dt}={4\over 3}{\sigma_T 
a_BT_0^4(1+z)^4\over m_p c}x_e.
\ee 
The product of the expansion rate $H$ and the velocity dissipation time $t_s$ is
\be
t_sH=7.6\times 10^{5}\,h\,x_e^{-1} (1+z)^{-5/2}.
\ee
Prior to decoupling $t_sH\ll 1$. Since the characteristic
time for the gravitational growth of mass density fluctuations is of the order of the
expansion timescale,  baryonic density fluctuations become free to grow only after decoupling.

We can use the above results to find the rate of relaxation of the matter
temperature $T_e$ to that of the radiation. The mean energy per electron in
the plasma is $E=3k_BT_e/2=m_e\langle v^2\rangle$. The rate at which an electron is doing
work against the radiation drag force if $Fv$, so the plasma transfers energy
to the radiation at the mean rate, per electron:
\be
-{dE\over dt}=\langle Fv\rangle={4\over 3}\,{\sigma_T a_BT^4\over c}\langle
v^2\rangle\propto T_e.
\ee
At thermal equilibrium $T_e=T$, and this rate must be balanced by the rate
at which photons scattering off electrons increase the matter energy:
\be 
{dE\over dt}={4\over 3}\,{\sigma_T a_BT^4\over c} {3k_B\over m_e} (T-T_e).
\ee 
Thus the rate of change of the matter temperature is
\be 
{dT_e\over dt}={2\over 3k_B}{dE\over dt}={x_e\over (1+x_e)}\, {8\sigma_T
a_BT^4\over 3m_ec}\,(T-T_e).
\ee 
The factor $x_e/(1+x_e)$ accounts for the fact that the plasma energy loss rate
per unit volume is $-n_e dE/dt=-x_en_\nH dE/dt$, while the total plasma energy
density is $(n_e+n_p+n_\nHI)3k_BT_e/2=n_\nH(1+x_e)3k_BT_e/2$. The expression above
can be rewritten as
\be 
{dT_e\over dt}={T-T_e\over t_c},
\ee 
where the ``Compton cooling timescale'' is
\be 
t_c={3m_ec\over 4\sigma_Ta_BT^4}\,{1+x_e\over 2x_e}={7.4\times 10^{19}~{\rm
s}\over (1+z)^4}\,\left({1+x_e\over 2x_e}\right).
\ee 
The characteristic condition for thermal coupling is then 
\be 
t_cH={240h\sqrt{\Omega_m}\over (1+z)^{5/2}}\,{1+x_e\over 2x_e}\,
\left({T\over T_e}-1\right)^{-1}<1.
\ee 
For fully ionized gas $x_e=1$ and the Compton cooling timescale is shorter
than the expansion time at all redshifts $z>z_c=5$. With
increasing redshift above 5, it becomes increasingly difficult
to keep optically thin ionized plasma hotter than the CMB.
At redshift $z>100$, well before the first energy sources (stars, accreting black 
holes) turn on, hydrogen will have the residual ionization $x_e=2.5\times 10^{-4}$. 
With this ionization, the characteristic relation for the relaxation of the matter temperature is
\be 
t_cH={1.7\times 10^5\over (1+z)^{5/2}}\,\left({T\over T_e}-1
\right)^{-1}.
\ee 
The coefficient of the fractional temperature difference reaches unity
at the ``thermalization redshift'' $z_{\rm th}\approx 130$. That is, the 
residual ionization is enough to keep the matter in
temperature equilibrium with the CMB well after decoupling. 
At redshift lower than $z_{\rm th}$ the temperature of intergalactic gas falls
adiabatically faster than that of the radiation, $T_e\propto a^{-2}$.
\begin{figure}
\vspace{0.cm}
\centerline{\psfig{figure=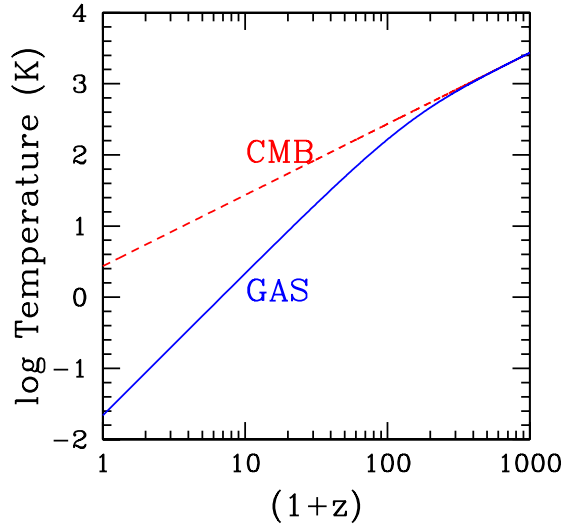,width=0.6\textwidth}}
\vspace{0.0cm}
\caption{Evolution of the radiation ({\it dashed line}, labeled CMB) and 
matter ({\it solid line}, labeled GAS) temperatures after recombination, in the
absence of any reheating mechanism.
}
\vspace{0.5cm}
\end{figure}
From the analysis above, the rate of change of the
radiation energy density due to Compton scattering can be written as
\be 
{du\over dt}={4\over 3}\,{\sigma_T a_BT^4\over c} {3k_Bn_e\over m_e} (T_e-T),
\ee 
or
\be 
{du\over u}=4dy,~~~~~~~dy\equiv (n_e\sigma_Tcdt)\,{k_B(T_e-T)\over m_ec^2}=d\tau_e
{k_B(T_e-T)\over m_ec^2}.
\ee 
Compton scattering causes a distorsion of the CMB spectrum, depopulating
the Rayleigh-Jeans regime in favor of photons in the Wien tail.
The ``Compton-parameter''
\be
y=\int_0^z {k_BT_e\over m_ec^2} {d\tau_e\over dz}dz
\ee 
is a dimensionless
measure of this distorsion, and is proportional to the pressure of the
electron gas $n_e k_BT_e$. The {\it COBE} satellite has shown the CMB to be
thermal to high accuracy, setting a limit $y\le 1.5\times 10^{-5}$ (Fixsen
et al. 1996). This can be shown to imply
\be 
\langle x_eT_e\rangle [(1+z)^{3/2}-1] <4\times 10^7\,{\rm K}.
\ee 
A universe that was reionized and reheated at $z=20$ to $(x_e, T_e)=(1, >4\times
10^5\,$K), for example, would violate the {\it COBE} $y$-limit.

\begin{table*}[t]
\caption{\sc Reaction network for primordial \HH\ chemistry}
\vspace{1em}
\begin{flushleft}
\begin{tabular}{llll}
\hline
   &  reaction   &   rate (cm$^3$~s$^{-1}$ or s$^{-1}$)   & notes \\
\hline
   &             &                            &                  \\
& {\bf H$^+$ + e $\rightarrow$ H + $\gamma$} & $R_{c2}=8.76\times 10^{-11}(1+z)^{-0.58}$  & \\
& {\bf H + $\gamma$ $\rightarrow$ H$^+$+ e} &  $R_{2c}=2.41\times 10^{15}\tr^{1.5}\exp{\left(-
\frac{39472}{\tr}\right)}\times R_{c2}$ & \\
& {\bf H + e $\rightarrow$ H$^-$ + $\gamma$}
   & $1.4\times 10^{-18}\tg^{0.928}\exp{\left(-\frac{\tg}{16200}\right)}$
   & 
   \\
& {\bf H$^-$ + $\gamma$ $\rightarrow$ H + e}
   & $1.1\times 10^{-1}\tr^{2.13}\exp{\left(-\frac{8823}{\tr}\right)}$
   &  
   \\
& {\bf H$^-$ + H $\rightarrow$ H$_2$ + e}
   & $1.5\times 10^{-9}$
   & $\tg\leq 300$
   \\
   &
   & $4.0\times 10^{-9}\tg^{-0.17}$
   & $\tg > 300$
   \\
& {\bf H$^-$ + H$^+$ $\rightarrow$ 2H }
   & $5.7\times 10^{-6}\tg^{-0.5}+6.3\times 10^{-8}-$
   &
   \\
   &
   & $9.2\times 10^{-11}\tg^{0.5}+4.4\times 10^{-13}\tg$
   &  
\\
& {\bf H + H$^+$ $\rightarrow$ H$_2^+$ + $\gamma$}
   & ${\rm dex}[-19.38-1.523\log\tg+$ 
   &
   \\
   &
   & $1.118(\log\tg)^2-0.1269(\log\tg)^3]$
   & $\tg\leq 10^{4.5}$
   \\
& {\bf H$_2^+$ + $\gamma$ $\rightarrow$ H + H$^+$}
   & $2.0\times 10^1\tr^{1.59}\exp{\left(-\frac{82000}{\tr}\right)}$
   & $v=0$
    \\
   &
   & $1.63\times 10^7\exp{\left(-\frac{32400}{\tr}\right)}$
   & LTE
   \\
& {\bf H$_2^+$ + H $\rightarrow$ H$_2$ + H$^+$}
   & $6.4\times 10^{-10}$
   &
   \\
& {\bf H$_2$ + H$^+$ $\rightarrow$ H$_2^+$ + H}
   & $3.0\times 10^{-10}\exp{\left(-\frac{21050}{\tg}\right)}$
   & $\tg\leq 10^4$
   \\
   &
   & $1.5\times 10^{-10}\exp{\left(-\frac{14000}{\tg}\right)}$
   & $\tg> 10^4$
   \\
& {\bf He + H$^+$ $\rightarrow$ HeH$^+$ + $\gamma$}
   & $7.6\times 10^{-18}\tg^{-0.5}$
   & $\tg\leq 10^3$,
   \\
   &
   & $3.45\times 10^{-16}\tg^{-1.06}$
   & $\tg> 10^3$ 
   \\
& {\bf HeH$^+$ + H $\rightarrow$ He + H$_2^+$}
   & $9.1\times 10^{-10}$
   &
   \\
& {\bf HeH$^+$ + $\gamma$ $\rightarrow$ He + H$^+$}
   & $6.8\times 10^{-1}\tr^{1.5}\exp\left(-\frac{22750}{\tr}\right)$
   &
   \\
\hline	
\end{tabular}
\vspace{1em}
\end{flushleft}
\end{table*}

\subsection{Hydrogen molecules in the early universe}

In the absence of heavy metals or dust grains, the processes of radiative cooling, 
cloud collapse, and star formation in the earliest astrophysical objects were
undoubtedly quite different from those today. Saslaw \& Zipoy (1967) and Peebles
\& Dicke (1968) were the first to realize the importance of gas phase
reactions for the formation of the simplest molecule, H$_2$, in the post-recombination
epoch. The presence of even a trace abundance of H$_2$ is of direct relevance
for the cooling properties of primordial gas. In the absence of molecules
this would be an extremely poor radiator: cooling by \Lya photons is 
ineffective at temperatures $\lta 8000$~K, well above the matter
and radiation temperature in the post-recombination era. It is the ability of
dust-free gas to cool down to low temperatures that controls 
the formation of the first stars in subgalactic systems.

The formation of H$_2$ in the intergalactic medium is 
catalyzed by the residual free electrons and ions, through the 
reactions  
\begin{eqnarray}
\rm
H \  \ \ \  + \ \ e^-  \  & \rightarrow & \ \ {\rm \mHm} \ \  +  \ \ \gamma,  \\
\rm \mHm  \ \ + \ \ H \ \ & \rightarrow & \ \ \rm \mHH  \ \ \  +  \ \  e^-, 
\end{eqnarray}
and
\begin{eqnarray}
\rm
\mHp \ \ + \ \ H  \ \ & \rightarrow & \ \ {\rm \mH2p} \ \  +  \ \ \gamma,  \\
\rm \mH2p  \ \ + \ \ H \ \ & \rightarrow & \ \ \rm \mHH   \ \  +  \ \ \mHp.
\end{eqnarray}
The second sequence produces molecular hydrogen at a higher redshift than the 
first, as the dissociation energy thresholds of the intermediaries species
H$_2^+$ and H$^-$ are 2.64 and 0.75 eV, respectively. The direct radiative 
association of H $+$ H $\rightarrow$ \HH\ $+$ $\gamma$ is highly forbidden, owing to the
negligible dipole moment of the homonuclear \HH\ molecule. The amount of \HH\ that forms
depends on the evolution of the gas density. Galli \& Palla (1998) have
analyzed in detail the case where the hydrogen follows the general expansion of the 
universe. Table 1 shows the reaction rates for hydrogen and helium species that enter in what 
they call the {\it minimal model}, i.e. the reduced set of processes that 
reproduces with excellent accuracy the full chemistry of \HH\ molecules. 
Column 1 gives the reaction, column 2 the rate coefficient 
(in cm$^3$~s$^{-1}$ for collisional processes, in s$^{-1}$ for photo-processes),
and column 3 the temperature range of validity of the rate coefficients and 
remarks on the rate (see Galli \& Palla 1998). The gas and radiation temperatures 
are indicated by $\tg$ and $\tr$, respectively.  

The evolution of the abundance of \HH\ follows the well known
behaviour (e.g. Lepp \& Shull 1984) where an initial steep
rise at $z\sim 400$ is determined by the H$_2^+$ channel, followed by a small contribution 
from H$^-$ at $z\sim 100$. The freeze-out primordial fraction of H$_2$ is [H$_2$/H] $\sim 
10^{-6}$. This is too small to trigger runaway collapse, fragmentation, and star formation.
The fate of a subgalactic system at high redshifts depends therefore on its 
ability to rapidly increase its \HH\ content following virialization.
In the low-density limit ($n_{\rm H}<0.1$ cm$^{-3}$), the \HH\ cooling 
function $\Lambda_{{\rm H}_2}$ 
(in erg~cm$^3$~s$^{-1}$) is well approximated by the expression
\begin{eqnarray}
\log \Lambda_{{\rm H}_2}[n_\nH\rightarrow 0] & = 
-103.0+97.59\log\tg-48.05(\log\tg)^2    \nonumber \\
& + 10.80(\log\tg)^3 -0.9032(\log\tg)^4,
\end{eqnarray}
over the range $10\;{\rm K}\leq \tg\leq 10^4\;{\rm K}$ (Galli \& Palla 1998).

\section{The Emergence of Cosmic Structure}

\subsection{Linear theory}

As shown above, it is only after hydrogen recombination that baryons can start falling 
into the already growing dark matter perturbations. Gas pressure can resist the force of gravity,
and small-scale perturbations in the baryonic fluid will not grow in amplitude. 
However, at sufficiently large scales, gravity can overpower pressure gradients, 
thereby allowing the perturbation to grow.

The linear evolution of sub-horizon density perturbations in the dark matter-baryon fluid is 
governed in the matter-dominated era by two second-order differential equations:
\be
\ddot \delta_{\rm dm}+2H\dot\delta_{\rm dm}\,=\, {3\over 2}
H^2 \Omega_m^z\,(f_{\rm dm}\delta_{\rm dm}+f_b\delta_b)
\label{eqlinear1}
\ee
for the dark matter, and
\be
\ddot \delta_b+2H\dot\delta_b \,=\, {3\over 2}
H^2 \Omega_m^z\,(f_{\rm dm}\delta_{\rm dm}+f_b\delta_b)\, -\, {c_s^2\over a^2} k^2
\delta_b
\label{eqlinear2}
\ee
for the baryons, where $\delta_{\rm dm}(k)$ and $\delta_b(k)$ are the Fourier
components of the density fluctuations in the dark matter and baryons,\footnote{For
each fluid component ($i=b,{\rm dm}$) the real space fluctuation in the density
field, $\delta_i({\bf x})\equiv \delta\rho_i({\bf x})/\rho_i$, can be written as a
sum over Fourier modes, $\delta_i({\bf x})=\int d^3{\bf k}\,(2\pi)^{-3}\,\delta_i({\bf k})
\,\exp{i{\bf k \cdot}{\bf x}}$.}\ $f_{\rm dm}$ and 
$f_b$ are the corresponding mass fractions, $c_s$ is the gas sound speed, $k$ 
the (comoving) wavenumber, and the derivates are taken with respect to cosmic time. Here
$\Omega_m^z\equiv 8\pi G \rho(t)/3H^2=\Omega_m(1+z)^3/[\Omega_m(1+z)^3+\Omega_\Lambda]$  
is the time-dependent matter density parameter, and $\rho(t)$ is the total background
matter density. Because there is 5 times more dark matter 
than baryons, it is the former that defines the pattern of gravitational wells in which
structure formation occurs. In the case where $f_b\simeq 0$ and the universe is static 
($H=0$), equation (\ref{eqlinear1}) becomes
\be
\ddot\delta_{\rm dm}=4\pi G\rho\delta_{\rm dm}\equiv {\delta_{\rm dm}
\over t_{\rm dyn}^2}, 
\ee
where $t_{\rm dyn}$ denotes the dynamical timescale. This equation admits solution 
$\delta_{\rm dm}=A_1\exp(t/t_{\rm dyn})+A_2\exp(-t/t_{\rm dyn})$.
After a few dynamical times, only the exponentially growing term is
significant: gravity tends to make small density fluctuations
in a static pressureless medium grow exponentially with time. The additional term
$\propto H\dot \delta_{\rm dm}$ present in an expanding universe can be thought
as a ``Hubble friction" term that acts to slow down the growth of density perturbations.
Equation (\ref{eqlinear1}) admits the general solution for the growing mode:
\be
\delta_{\rm dm}(a)={5\Omega_m\over 2}~{H_0^2}\,H\int_0^a {da'\over  (\dot a')^3},
\label{growth}
\ee
where the constant have been choosen so that an Einstein-de Sitter universe 
($\Omega_m=1$, $\Omega_\Lambda=0$) gives the familiar scaling $\delta_{\rm dm}(a)=a$ with coefficient 
unity. The right-hand side of equation (\ref{growth}) is called the linear growth 
factor $D(a)$. Different
values of $\Omega_m,\Omega_\Lambda$ lead to different linear growth factors:
growing modes actually decrease in density, but not 
as fast as the average universe. Note how, in contrast to the exponential growth found 
in the static case, the growth of perturbations even in the case of an Einstein-de
Sitter universe is just algebraic. A remarkable approximation formula to the growth factor
in a flat universe follows from Lahav et al. (1991),  
\be
\delta_{\rm dm}(a)=D(a)\simeq {5\Omega_m^z\over 2(1+z)}\left[({\Omega_m^z})^{4/7}-
{(\Omega_m^z)^2\over 140}+{209\over 140}\Omega_m^z+{1\over 70}\right]^{-1}.
\label{growthL}
\ee
This is good to a few percent in regions of plausible $\Omega_m, \Omega_\Lambda$.

\begin{figure}
\begin{center}
\includegraphics[width=.49\textwidth]{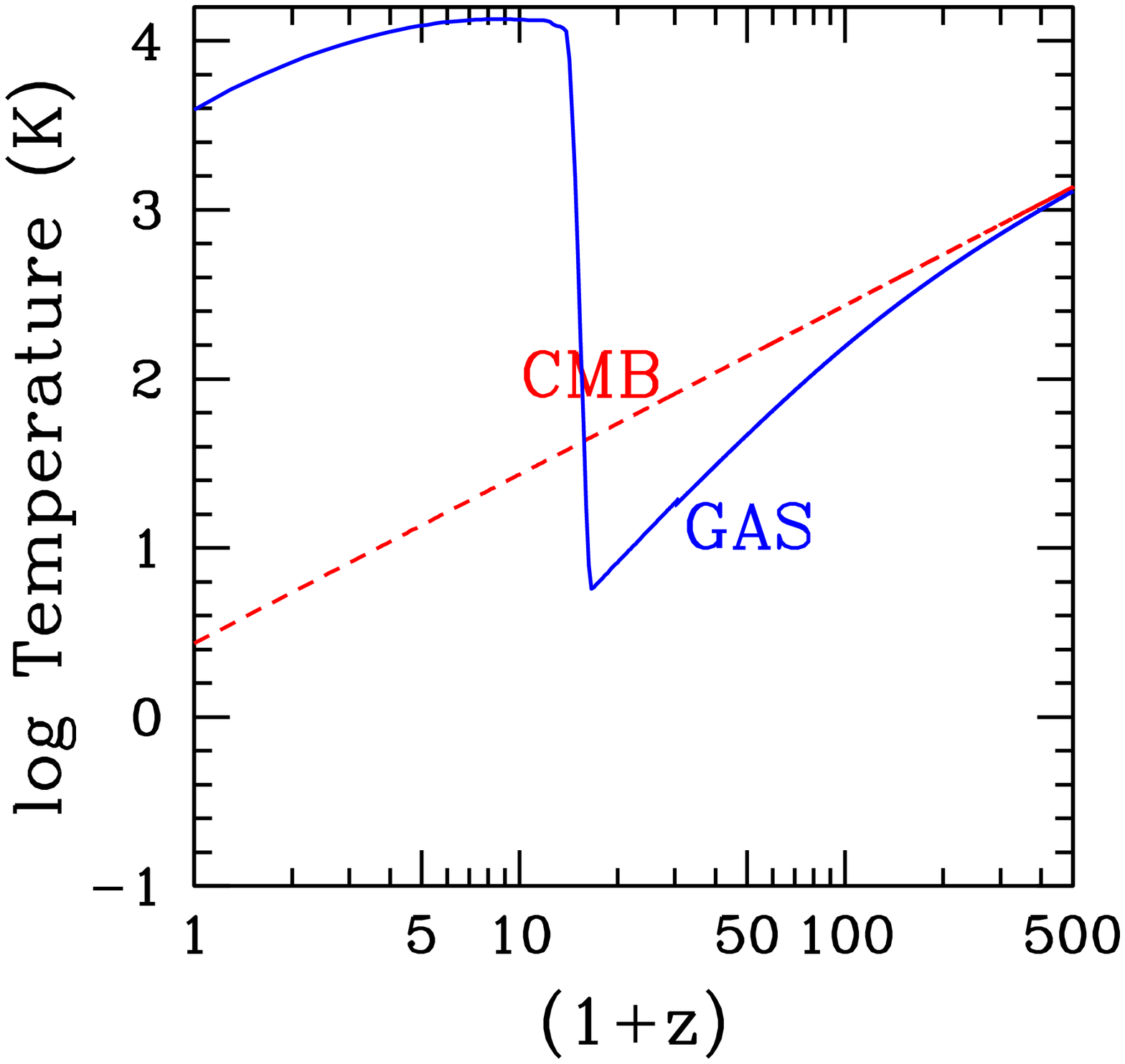}
\includegraphics[width=.49\textwidth]{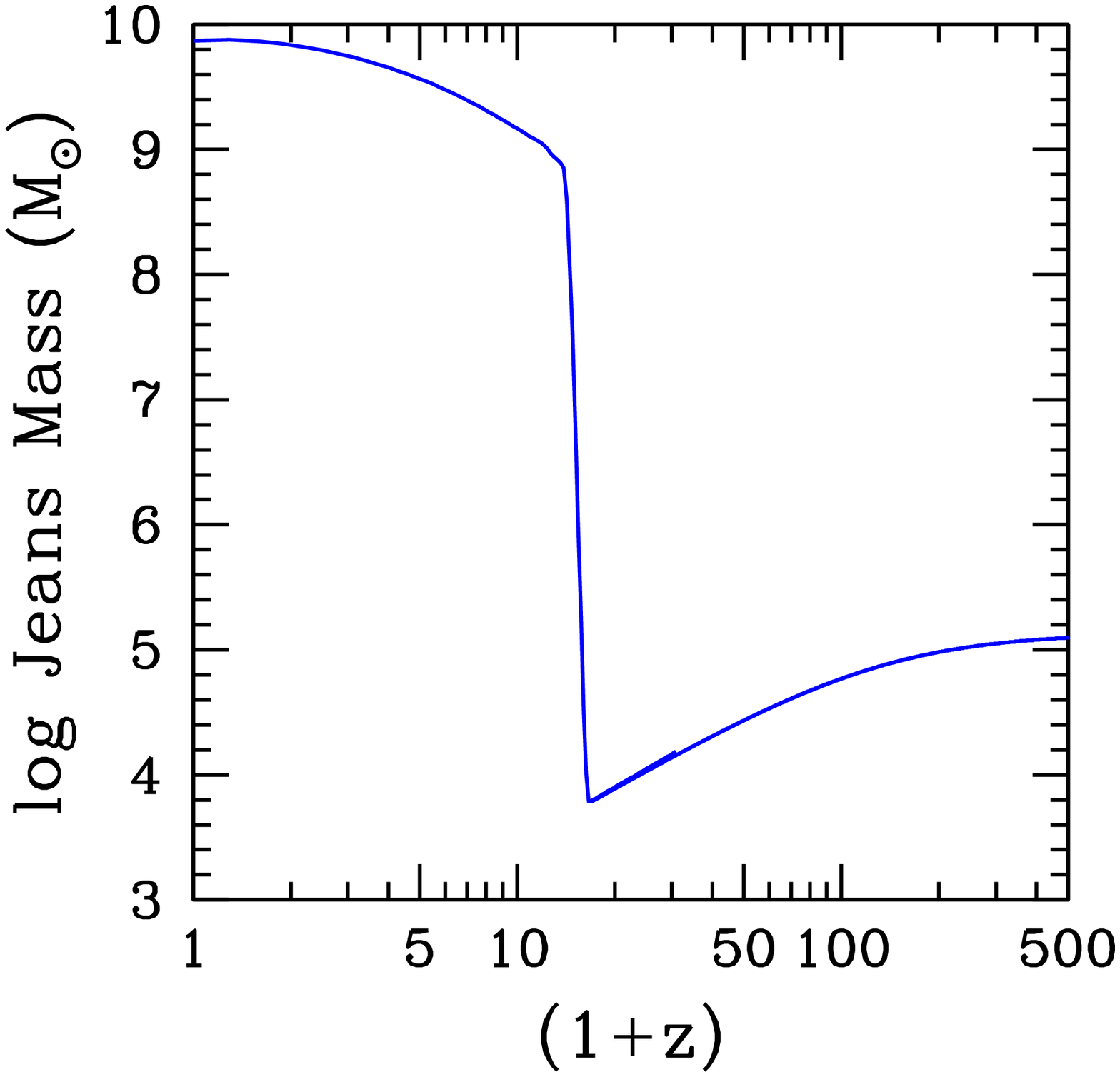}
\end{center}
\caption[]{ \textit{Left:} Same as Fig. 3 but for a universe 
reionized by ultraviolet radiation at $z_r=11$. 
\textit{Right:} Cosmological (gas $+$ dark matter) Jeans mass.
\label{fig:Tevolution}
}
\end{figure}

Equation (\ref{eqlinear2}) shows that, on large scales (i.e. small $k$), pressure forces
can be neglected and baryons simply track the dark matter fluctuations. On the contrary, on small
scales (i.e. large $k$), pressure dominates and baryon fluctuations will be suppressed
relative to the dark matter. Gravity and pressure are equal at the characteristic Jeans
scale:
\be 
k_J={a\over c_s} \sqrt{4\pi G \rho}.
\label{eqjeanswave}
\ee 
Corresponding to this critical wavenumber $k_J$ there is a critical 
cosmological Jeans mass $M_J$, defined as the total (gas $+$ dark 
matter) mass enclosed within 
the sphere of physical radius equal to $\pi a/k_J$,
\be
M_J={4\pi\over 3}\rho \left({\pi a\over k_J}\right)^3=
{4\pi\over 3}\rho\left(\frac{5\pi k_BT_e}{12 G\rho m_p\mu}\right)^{3/2}
\approx 8.8\times 10^4\,\msun \left({aT_e\over \mu}\right)^{3/2}, 
\ee
where $\mu$ is the mean molecular weight. The evolution of $M_J$ is shown in
Figure~\ref{fig:Tevolution}. In the post-recombination universe, the
baryon-electron gas is thermally coupled to the CMB, $T_e\propto
a^{-1}$, and the Jeans mass is independent of redshift and comparable
to the mass of globular clusters, $M_J\approx 10^5\,\msun$.  For
$z<z_{\rm th}$, the gas temperature drops as $T_e \propto a^{-2}$,
and the Jeans mass decreases with time, $M_J\propto a^{-3/2}$.  This
trend is reversed by the reheating of the IGM. The energy released by
the first collapsed objects drives the Jeans mass up to galaxy scales
(Figure~\ref{fig:Tevolution}): baryonic density perturbations stop
growing as their mass drops below the new Jeans mass. In
particular, photo-ionization by the ultraviolet radiation from the
first stars and quasars would heat the IGM to temperatures of $\approx
10^4\,$K (corresponding to a Jeans mass $M_J \sim 10^{10}\,\msun$
at $z_r=11$), suppressing gas infall into low mass systems and
preventing new (dwarf) galaxies from forming.\footnote{When the Jeans 
mass itself varies with time, linear gas fluctuations
tend to be smoothed on a (filtering) scale that depends on the full
thermal history of the gas instead of the instantaneous value of the
sound speed (Gnedin \& Hui 1998).}  

\subsection{Statistics of density fields}

The observed uniformity of the CMB guarantees that density fluctuations 
must have been quite small at decoupling, implying that the 
evolution of the density contrast can be studied at $z\lta z_{\rm dec}$ 
using linear theory, and each mode $\delta(k)$ evolves independently. 
The inflationary model predicts a scale-invariant primordial power spectrum 
of density fluctuations $P(k)\equiv \langle |\delta(k)|^2\rangle \propto
k^n$, with $n=1$ (the so-called Harrison-Zel'dovich spectrum). It is the index 
$n$ that governs the balance between large and small-scale
power. In the case of a Gaussian random field with zero mean, the power spectrum 
contains the complete statistical information about the density inhomogeneity.
It is often more convenient to use the dimensionless quantity
$\Delta^2_k\equiv [k^3P(k)/2\pi^2]$, which is the power per logarithmic interval 
in wavenumber $k$. In the matter-dominated epoch, this
quantity retains its initial primordial shape ($\Delta^2_k\propto k^{n+3}$)
only on very large scales. Small wavelength modes enter the horizon
earlier on and their growth is suppressed more severely during the radiation-dominated epoch:
on small scales the amplitude of $\Delta^2_k$ is essentially suppressed
by four powers of $k$ (from $k^{n+3}$ to $k^{n-1}$). If $n=1$, then small scales
will have nearly the same power except for a weak, logarithmic dependence.
Departures from the initially scale-free form  are described by the transfer
function $T(k)$, defined such that $T(0)=1$:
\be
P(k,z)=Ak^n\left[{D(z)\over D(0)}\right]^2T^2(k),
\ee
where $A$ is the normalization. An accurate fitting function for $T(k)$ 
in a $\Lambda$CDM universe is
\be
T_k={\ln(1+2.34q)\over 2.34 q}\left[1+3.89q+(16.1q)^2 +(5.46q)^3
  +(6.71q)^4\right]^{-1/4},
\ee
where 
\be
q\equiv {k/{\rm Mpc^{-1}}\over \Omega_m h^2 \exp(-\Omega_b-\Omega_b/\Omega_m)}.
\ee
This is the fit given by Bardeen \etal (1986) modified to account for the effects 
of baryon density following Sugiyama (1995).

Another useful measurement of inhomogeneity is the quantity
\be
\sigma^2(M)={1\over 2\pi^2}\int_0^\infty dkk^2P(k)\,|W(kR)|^2.
\ee
This is the mass variance of the density field smoothed by a window function 
$W(kR)$ over a spherical volume of comoving radius $R$ and average mass  
$M=H_0^2\Omega_mR^3/(2G)$. If the window is a top-hat in real space, then
its Fourier transform is $W(kR)=3(kR)^{-3}\,(\sin kR -kR\cos kR)$.
The significance of the window function is the following: 
the dominant contribution to $\sigma(M)$ comes from perturbation components
with wavelengths $\lambda=2\pi/k>R$, because those with higher frequencies 
tend to be averaged out within the window volume. As the fluctuation spectrum 
is falling with decreasing $k$, waves with much larger $\lambda$ contribute 
only a small amount.  Hence, in terms of a  mass $M\propto \lambda^3\propto k^{-3}$,
we have
\be
\sigma(M)\propto M^{-(3+n)/n}.
\ee
Since $P(k)$ is
not a strict power law, $n$ should be thought of as an approximate local value $d\ln P/
d\ln k$ in the relevant range. For $n>-3$, the variance $\sigma_M$ decreases with
increasing $M$; this implies that, on the average, smaller masses condense out
earlier than larger masses.  Structures grow by the gradual
separation and collapse of progressively larger units. Each parent unit will in general 
be made up of a number of smaller progenitor clumps that had collapsed earlier. 
This leads to a  hierarchical pattern of clustering.
In CDM, $n\ge -3$ at small $M$, increases with increasing $M$ and reaches
the asymptotic value $n=1$ for $M\gta 10^{15}\,\msun$. The power spectrum on 
galactic scales can be approximated by $n\approx -2$. 

\subsection{Spherical collapse} 

At late times the density contrast at a given
wavelength becomes comparable to unity, linear perturbation theory 
fails at this wavelength, and a different model must be used to follow
the collapse of bound dark matter systems (``halos"). Small scales are the first 
to become non-linear. 
Consider, at some initial time $t_i$, a spherical region of size $r_i$
and mass $M$ that has a slight constant overdensity $\delta_i$ relative to the 
background $\rho_b$, 
\be
\rho(t_i)=\rho_b(t_i)(1+\delta_i).
\ee
As the universe expands, the overdense region will expand 
more slowly compared to the background, will reach a maximum radius, and 
eventually collapse under its own gravity to form a bound virialized system.
Such a simple model is called a spherical top-hat. As long as radial shells
do not cross each other during the evolution, the motion of a test particle
at (physical) radius $r$ is governed by the equation (ignoring the vacuum energy component)
\be
{d^2r\over dt^2}=-{GM\over r^2}, 
\label{eqmotion}
\ee
where $M=(4\pi/3)r_i^3\rho_b(t_i)(1+\delta_i)=$const.
Integrating we obtain 
\be
{1\over 2}\left({dr\over dt}\right)^2-{GM\over r}=E, 
\ee
where $E$ is a constant of integration. If $E>0$ then $\dot r^2$ will never
become zero and the shell will expand forever. If $E<0$ instead, then as 
$r$ increases $\dot r$ will eventually become zero and later negative,      
implying a contraction and a collapse. Let's choose $t_i$ to be the time
at which $\delta_i$ is so small that  the overdense region is expanding
along with the Hubble flow. Then $\dot r_i=(\dot a/a)r_i=H(t_i)r_i
\equiv H_ir_i$ at time $t_i$, and the initial kinetic energy will be
\be
K_i\equiv \left({\dot r^2\over 2}\right)_{t=t_i}={H_i^2r_i^2\over 2}.  
\ee
The potential energy at $t=t_i$ is 
\be
|U|=\left({GM\over r}\right)_{t=t_i}=G{4\pi\over 3}\rho_b(t_i)(1+\delta_i)
r_i^2={1\over 2}H_i^2r_i^2\Omega_i(1+\delta_i)=K_i\Omega_i(1+\delta_i),
\ee
with $\Omega_i$ denoting the initial value of the matter density parameter
of the smooth background universe. The total energy of the shell is 
therefore
\be
E=K_i-K_i\Omega_i(1+\delta_i)=K_i\Omega_i(\Omega_i^{-1}-1-\delta_i). 
\ee
The condition $E<0$ for the shell to eventually collapse becomes
$(1+\delta_i)>\Omega_i^{-1}$, or
\be
\delta_i>\Omega_i^{-1}-1.
\ee
In a Einstein-de Sitter universe at early times ($\Omega_i=1$), this condition is
satisfied by any overdense region with $\delta_i>0$. In this case the patch
will always collapse.

Consider now a shell with $E<0$ in a background Einstein-de Sitter universe
that expands to a maximum radius $r_{\rm max}$ (`turnaround', $\dot r=0$) and then 
collapses. The solution 
to the equation of motion can be written as the parametric equation for a cycloid
\be
r={r_{\rm max}\over 2}(1-\cos\theta),~~~~~t=t_{\rm max}\,{\theta-\sin\theta
\over \pi}. 
\ee
The background density evolves as $\rho_b(t)=(6\pi Gt^2)^{-1}$, and the density contrast 
becomes 
\be
\delta={\rho(t)\over \rho_b(t)}-1={9\over 2}{(\theta-\sin\theta)^2\over (1-\cos\theta)^3}-1.
\ee
For comparison, linear theory (in the limit of small $t$) gives
\be
\delta_L={3\over 20}\left({6\pi t\over t_{\rm max}}\right)^{2/3}=
{3\over 20}[6(\theta-\sin\theta)]^{2/3}.
\ee 
This all agrees with what we knew already: at early times the sphere expands with 
the $a\propto t^{2/3}$ Hubble flow and density perturbations grow proportional to $a$.
We can now see how linear theory breaks down as the perturbation evolves.
There are three interesting epochs in the final stage of its development,
which we can read directly from the above solutions.

\bigskip\noindent
{\bf Turnaround.} The sphere breaks away from the Hubble expansion and 
reaches a maximum radius at $\theta=\pi$, $t=t_{\rm max}$. At this point the 
true density enhancement with respect to the background is 
\be
\delta={9\over 2}{\pi^2\over 2^3}-1={9\over 16}\pi^2-1=4.55,
\label{delta_max}
\ee
which is definitely in the non-linear regime. By comparison, linear theory predicts 
\be
\delta_L={3\over 20}(6\pi)^{2/3}=1.062.
\ee

\bigskip\noindent
{\bf Collapse.} If only gravity operates, then the sphere will collapse to a
singularity at $\theta=2\pi$, $t=2t_{\rm max}$. This occurs when
\be
\delta_L={3\over 20}(12\pi)^{2/3}=1.686.
\ee
Thus from equation (\ref{growthL}) we see that a top-hat collapses at redshift 
$z$ if its linear overdensity {\it extrapolated to the present day} is
\be
\delta_c(z)=\frac{\delta_L}{D(z)}=1.686(1+z),
\label{deltac} 
\ee 
where the second equality holds in a flat Einstein-de Sitter universe. This is 
termed the critical overdensity for collapse. An object of mass $M$ collapsing at 
redshift $z$ has an overdensity that is $\nu_c$ times the linearly extrapolated density
contrast today, $\sigma_0(M)$, on that scale,  
\be
\nu_c=\delta_c(z)/\sigma_0(M).
\ee
The mass scale associated with typical $\nu_c=1$-$\sigma$ non-linear 
fluctuations in the density field decreases from 
$10^{13}\,\msun$ today to about $10^7\,\msun$ at redshift 5. 
At $z=10$, halos of $10^{10}\,\msun$ collapse from much rarer $\nu_c=3$-$\sigma$ 
fluctuations.  
 
\bigskip\noindent
{\bf Virialization.} Collapse to a point at $\theta=2\pi$ ($t=2t_{\rm max}
\equiv t_{\rm vir}$) will never occur in 
practice as, before this happens, the approximation that matter is
distributed in spherical shells and that the random velocities of the particles
are small will break down. The dark matter will reach virial equilibrium
by a process known as `violent relaxation': since the gravitational potential is 
changing with time, individual particles do not follow orbits that conserve 
energy, and the net effect is to widen the range of energies available to them.
Thus a time-varying potential can provide a relaxation mechanism that 
operates on the dynamical timescale rather than on the much longer two-body relaxation time.
This process will convert the kinetic energy of collapse into random motions.

At $t=t_{\rm max}$ all the energy is in the form of gravitational potential
energy, and $E=U=-GM/r_{\rm max}$. At virialization $U=-2K$ (virial 
theorem) and $E=U+K=U/2=-GM/(2r_{\rm vir})$. Hence $r_{\rm vir}=r_{\rm max}/2$. 
The mean density of the virialized object is then $\rho_{\rm vir}=2^3\,\rho_{\rm 
max}$, where $\rho_{\rm max}$ is the density of the shell at turnaround. 
From equation (\ref{delta_max}) we have $\rho_{\rm max}=(9/16)\pi^2\rho_b(t_{\rm max})$, and
$\rho_b(t_{\rm max})=\rho_b(t_{\rm vir})(t_{\rm vir}/t_{\rm max})^2=
4\,\rho_b(t_{\rm vir})$. Combining these relations we get:
\be
\rho_{\rm vir}=2^3\rho_{\rm max}=2^3(9/16)\pi^2\rho_b(t_{\rm max})=
2^3(9/16)\pi^2 4\,\rho_b(t_{\rm vir})=18\pi^2\,\rho_b(t_{\rm vir}).
\ee
Therefore the density contrast at virialization in an Einstein-de Sitter universe
is 
\be
\Delta_c=178.
\ee 

\medskip\noindent
In a universe with a cosmological constant, the collapse of a top-hat spherical
perturbation is described by
\be
{d^2r\over dt^2}=-{GM\over r^2}+{\Lambda\over 3}c. 
\label{eqmotionL}
\ee
In a flat $\Omega_m+\Omega_\Lambda=1$ cosmology, the final overdensity {\it relative to
the critical density} gets modified according to the fitting formula (Bryan 
\& Norman 1998) 
\be
\Delta_c=18\,\pi^2+82d-39d^2,
\ee
where $d\equiv \Omega_m^z-1$ is evaluated at the collapse redshift. 
A spherical top-hat collapsing today in a universe with $\Omega_m=0.24$ has a density
contrast at virialization of $\Delta_c=93$. This corresponds to an overdensity relative
to the background matter density ($=\Omega_m\rho_c^0$) of $\Delta_c/\Omega_m=388$: the faster 
expansion of a low-density universe means that the perturbation turns around and 
collapses when a larger density contrast has been produced. For practical reason a density
contrast of 200 relative to the background is often used to define the radius, $\rtwo$,
that marks the boundary of a virialized region.

\medskip\noindent
A halo of mass $M$ collapsing at redshift $z$ can be described in terms 
of its virial radius $r_{\rm vir}$, circular velocity $V_c$ and
virial temperature $T_{\rm vir}$ (Barkana \& Loeb 2001):
\be
r_{\rm vir}=\left[{2GM\over \Delta_c H^2}\right]^{1/3}=
1.23\,{\rm kpc}\,\left({M\over 10^8\,\msun}\right)^{1/3}\,
f^{-1/3}\,
\left({1+z\over 10}\right)^{-1},
\ee
\be
V_c=\left({GM\over r_{\rm vir}}\right)^{1/2}=
21.1\,{\rm km~s^{-1}}\,\left({M\over 10^8\,\msun}\right)^{1/3}\,
f^{1/6}\,
\left({1+z\over 10}\right)^{1/2}, 
\ee
\be
T_{\rm vir}={\mu m_p V_c^2\over 2k_B}=1.6\times 10^4\,{\rm K}\,
\left({M\over 10^8\,\msun}\right)^{2/3}\,
f^{1/3}\,
\left({1+z\over 10}\right), 
\ee
where $f\equiv (\Omega_m/\Omega_m^z)(\Delta_c/18\pi^2)$ and $\mu=0.59$ is the 
mean molecular weight for fully ionized primordial gas ($\mu=1.23$ for neutral primordial gas).
The binding energy of the halo is approximately
\be 
E_b={1\over 2} {GM^2\over r_{\rm vir}}=4.42\times 10^{53}\,{\rm erg}\,
\left({M\over 10^8\,\msun}\right)^{5/3}\,
f^{1/3}\,
\left({1+z\over 10}\right),  
\ee
where the coefficient of $1/2$ is exact for a singular isothermal sphere.
The binding energy of the baryons is smaller by a factor 
$\Omega_m/\Omega_b\simeq 5.7$. The energy deposition
by supernovae in the shallow potential wells of subgalactic systems 
may then lift out metal-enriched material from the host (dwarf) halos,
causing the pollution of the IGM at early times (Madau \etal 2001).

\begin{figure}
\centerline{\psfig{figure=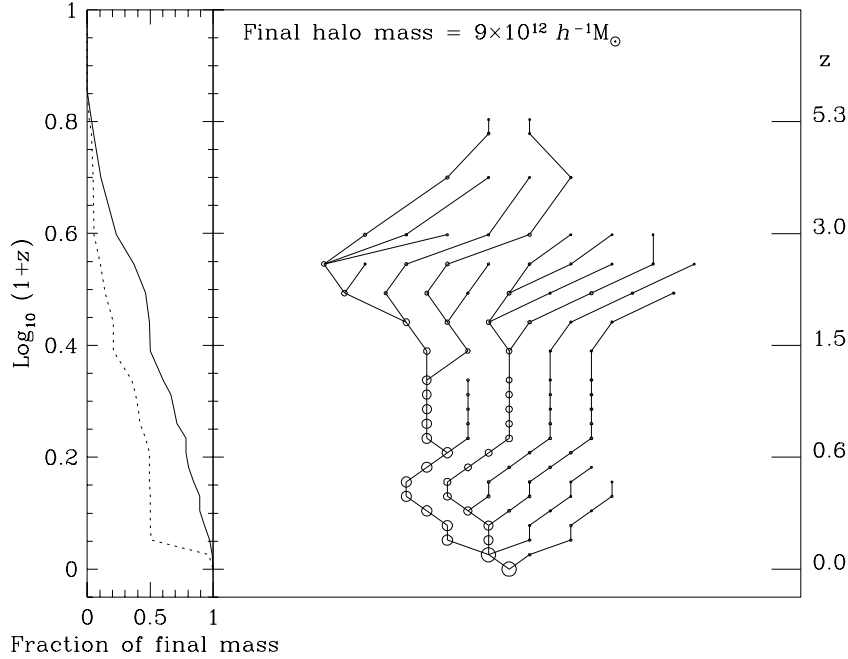,width=0.9\textwidth}}
\vspace{0.0cm}
\caption{An example of a merger tree obtained from an N-body simulation of a $9 \times 
10^{12}h^{-1}\,\msun$ halo at redshift $z=0$. Each circle represents a dark matter halo 
identified in the simulation, the area of the circle being proportional to halo mass. 
The vertical position of each halo on the plot is determined by the redshift $z$ at which 
the halo is identified, the horizontal positioning is arbitrary. The solid 
lines connect halos to their progenitors. The solid line in the panel on the 
left-hand side shows the fraction of the final mass contained in resolved progenitors as 
a function of redshift. The dotted line shows the fraction of the final mass 
contained in the largest progenitor as a function of redshift. (From Helly \etal 2003.)
\label{fig:Helly}
}
\end{figure}
\begin{figure}
\centerline{\psfig{figure=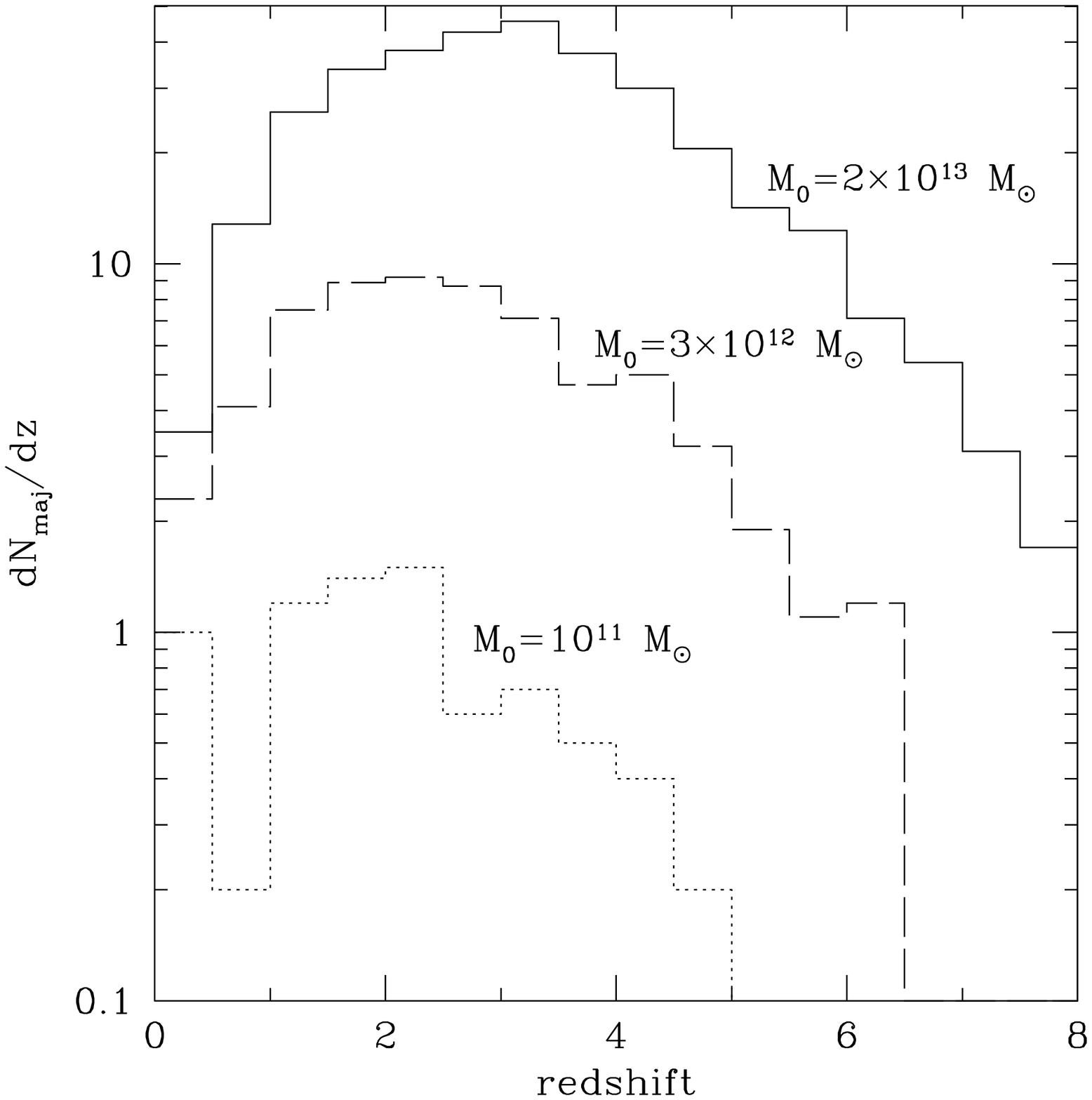,width=0.6\textwidth}}
\caption{Mean number of major mergers experienced per unit
redshift by halos with masses $>10^{10}\,\msun$. {\it Solid line:}
progenitors of a $M_0=2\times 10^{13}\,\msun$ halo at $z=0$. {\it Dashed line:}
same for $M_0=3\times 10^{12}\,\msun$. {\it Dotted line:} same for
$M_0=10^{11}\,\msun$. (From Volonteri \etal 2003.)}
\label{figmarta}
\end{figure}

\subsection{Dark halo mergers}

The assumption that virialized objects form from smooth spherical collapse,
while providing a useful framework for thinking about the formation histories 
of gravitationally bound dark matter halos, does not capture the real nature
of structure formation in CDM theories. In these models galaxies are assembled
hierarchically through the merging of many smaller subunits that
formed in a similar manner at higher redshift (see Fig. \ref{fig:Helly}). 
Galaxy halos experience multiple mergers during their lifetime, with those between 
comparable-mass systems (``major mergers'') expected to result in the formation 
of elliptical galaxies (see, e.g., Barnes 1988; Hernquist 1992). Figure \ref{figmarta} 
shows the number of major mergers (defined as mergers where the mass ratio of 
the progenitors is $>0.3$) per unit redshift bin experienced by halos of different 
masses. For galaxy-sized halo this quantity happens to peak in the redshift range 
2-4, the epoch when the observed space density of optically-selected quasar also 
reaches a maximum.

The merger between a large parent 
halo and a smaller satellite system will evolve under two dynamical processes:
dynamical friction, that causes the orbit of the satellite to decay toward the 
central regions, and tidal stripping, that removes material from the
satellite and adds it to the diffuse mass of the parent. Since clustering is hierarchical,
the satellite will typically form at earlier times and have a higher characteristic density 
and a smaller characteristic radius. The study of the assembly history of dark matter
halos by repeated mergers is particularly well suited to the N-body 
methods that have been developed in the past two decades. Numerical simulations
of structure formation by dissipationless hierarchical clustering from Gaussian 
initial conditions indicate a roughly universal spherically-averaged 
density profile for the resulting halos (Navarro \etal 1997, hereafter NFW):
\be
\rho_{\rm NFW}(r)=\frac{\rho_s}{cx(1+cx)^2}, 
\ee
where $x\equiv r/\rvir$ and the characteristic density $\rho_s$ is 
related to the concentration parameter $c$ by
\be
\rho_s={3H^2\over 8\pi G}\,{\Delta_c\over 3}\,{c^3\over \ln(1+c)-c/(1+c)}.
\ee 
This function fits the numerical data presented by NFW over a radius range of 
about two orders of magnitude. Equally good fits are obtained for high-mass
(rich galaxy cluster) and low-mass (dwarf) halos. Power-law fits to this profile
over a restricted radial range have slopes that steepen from $-1$ near the halo
center to $-3$ at large $cr/\rvir$. Bullock \etal (2001)
found that the concentration parameter follows a log-normal distribution
where the median depends on the halo mass and redshift,
\be
  c_{\rm med}(M,z) = \frac{\medc}{1+z} \left(\frac{M}{M_*}\right)^{\alpha} ,
\ee
where $M_*$ is the mass of a typical halo collapsing today.
The halos in the simulations by Bullock \etal were best
described by $\alpha=-0.13$ and $\medc=9.0$, with a scatter around the
median of $\sigc=0.14$~dex.
\begin{figure}
\centerline{\psfig{figure=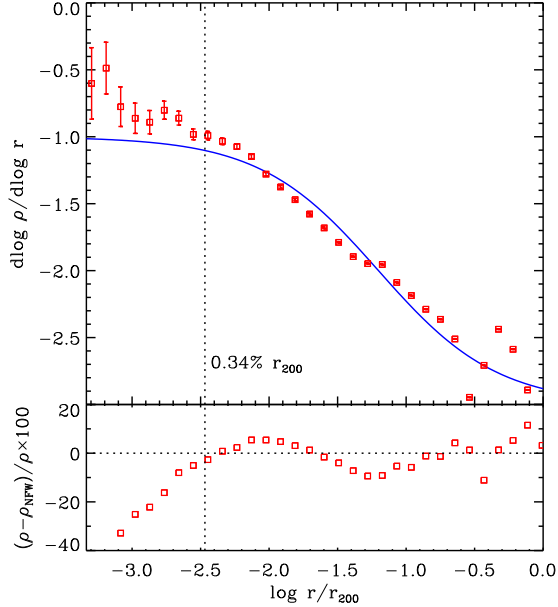,width=0.6\textwidth}}
\caption{\textit{Top}: logarithmic slope of the density profile of the
``Via Lactea'' halo, as a function of radius. Densities were computed in 50 radial
logarithmic bins, and the local slope was determined by a finite
difference approximation using one neighboring bin on each side. The
thin line shows the slope of the best-fit NFW profile with concentration
$c=12$. The vertical dotted line indicates the
estimated convergence radius: local densities
(but not necessarily the logarithmic slopes)
should be correct to within 10\% outside of this radius.
\textit{Bottom}: the residuals in
percent between the density profile and the best-fit NFW profile, as a
function of radius. Here $r_{200}$ is the radius within which the 
enclosed average density is 200 times the background value, $\rtwo=1.35\,\rvir$ 
in the adopted cosmology. (From Diemand \etal 2007a.)
}
\label{profile}
\end{figure}

``Via Lactea", the highest resolution N-body simulation to date of the formation of 
a Milky Way-sized halo (Diemand \etal 2007a,b), 
shows that the fitting formula proposed by NFW with concentration $c=12$ 
provides a reasonable approximation to the density profile down a convergence 
radius of $r_{\rm conv}=1.3\,$kpc.  Within the region of convergence, deviations from
the best-fit NFW matter density are typically less than 10\%. From 10 kpc down to
$r_{\rm conv}$ Via Lactea is actually denser than predicted by the NFW formula.
Near $r_{\rm conv}$ the density approaches the NFW value again while the logarithmic
slope is shallower ($-1.0$ at $r_{\rm conv}$) than predicted by the NFW fit.

\subsection{Assembly history of a Milky Way halo}

\begin{figure}
\vspace{0.3cm}
\centerline{\epsfig{figure=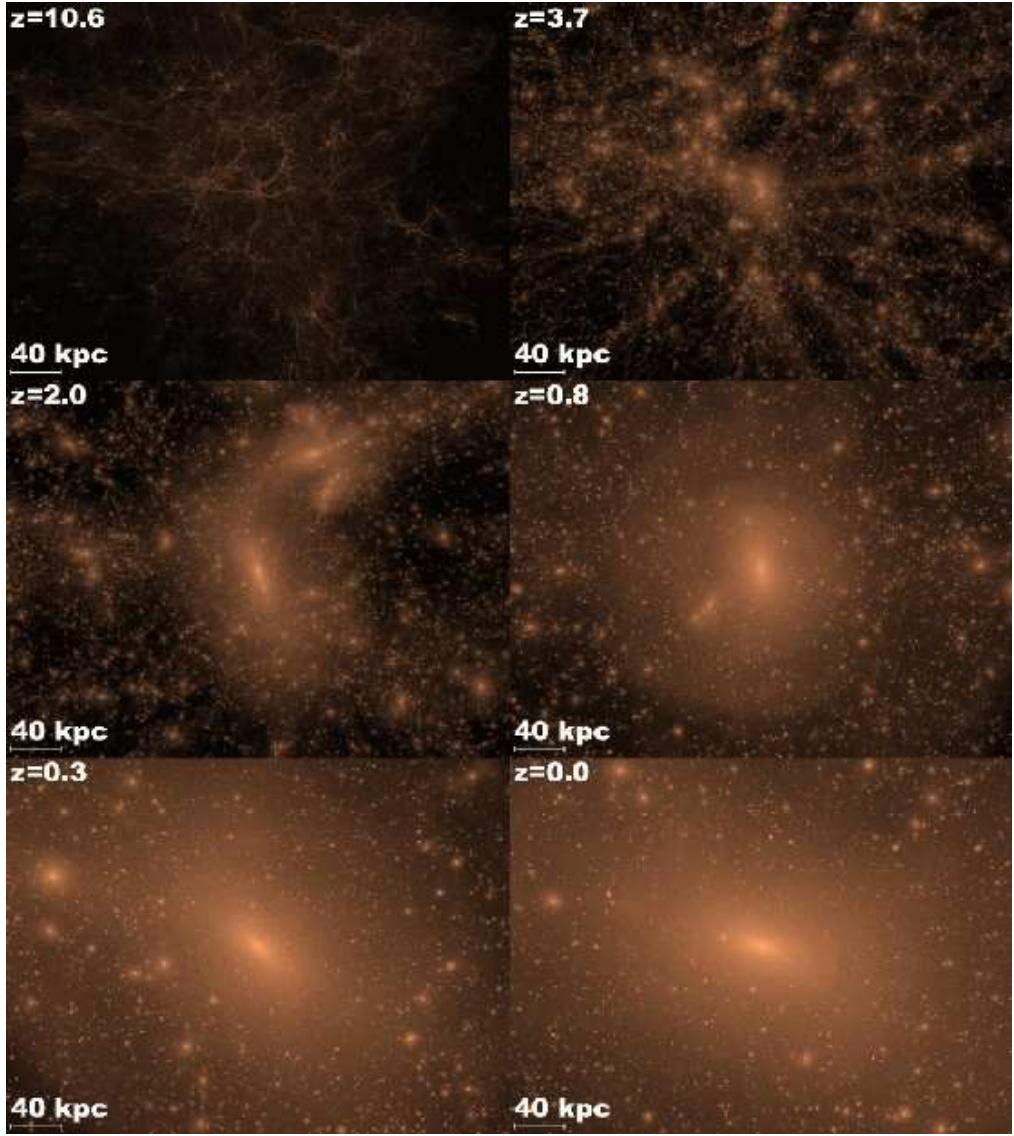,width=0.99\textwidth}}
\vspace{0.4cm}
\caption{Projected dark matter density-squared map of our simulated
Milky Way halo region (``Via Lactea'') at different redshift, from $z=10.6$ to the 
present. The image covers an area of 400 $\times$ 300 physical kpc.}
\label{L200}
\end{figure}

The simple spherical top-hat collapse ignores shell crossing and
mixing, accretion of self-bound clumps, triaxiality, angular momentum, 
random velocities, and large scale tidal forces. 
It is interesting at this stage to use ``Via Lactea"  and study in more details, starting from 
realistic initial conditions, the formation history of a Milky Way-sized halo in a $\Lambda$CDM
cosmology. The Via Lactea simulation was performed with the PKDGRAV tree-code
(Stadel 2001)  using the best-fit cosmological parameters from the
{\it WMAP} 3-year data release. The galaxy-forming region was sampled with
234 million particles of mass $2.1\times 10^4\msun$, evolved
from redshift 49 to the present with a force resolution of 90 pc and 
adaptive time-steps as short as 
68,500 yr, and centered on an isolated halo that had no major
merger after $z=1.7$, making it a suitable host for a Milky Way-like
disk galaxy (see Fig. \ref{L200}). The number of particles is an order of magnitude larger than used in previous
simulations. The run was completed in 320,000 CPU hours on NASA's
Project Columbia supercomputer, currently one of the fastest machines
available. (More details about the Via Lactea run are given in Diemand \etal 2007a,b.
Movies, images, and data are available at http://www.ucolick.org/$\sim$diemand/vl.) 
The host halo mass at $z=0$ is $\mtwo=1.8\times 10^{12}\,\msun$ within a radius of 
$\rtwo=389\,$ kpc. 
 
\begin{figure}
\centerline{\epsfig{figure=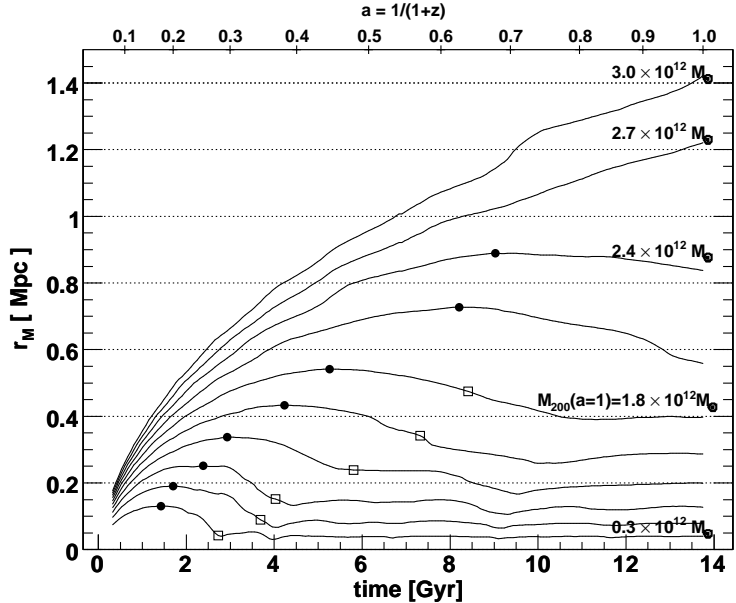,width=0.8\textwidth}}
\vspace{0.0cm}
\caption{Evolution of radii $r_M$ enclosing a fixed mass versus cosmic time or scale factor $a$. 
The enclosed mass grows in constant amounts of $0.3\times 10^{12}\,\msun$
from bottom to top. Initially all spheres are growing in the physical 
(non comoving) units used here. Inner shells turn around, collapse and stabilize, while
the outermost shells are still expanding today. {\it Solid circles}: points of 
maximum expansion at the turnaround time $t_{\rm max}$. {\it Open squares:} time after 
turnaround where $r_M$ first contracts within $20\%$ of the final value. These mark 
the approximate epoch of ``stabilization''.
}
\label{LagRadii}
\end{figure}

\begin{figure}
\centerline{\epsfig{figure=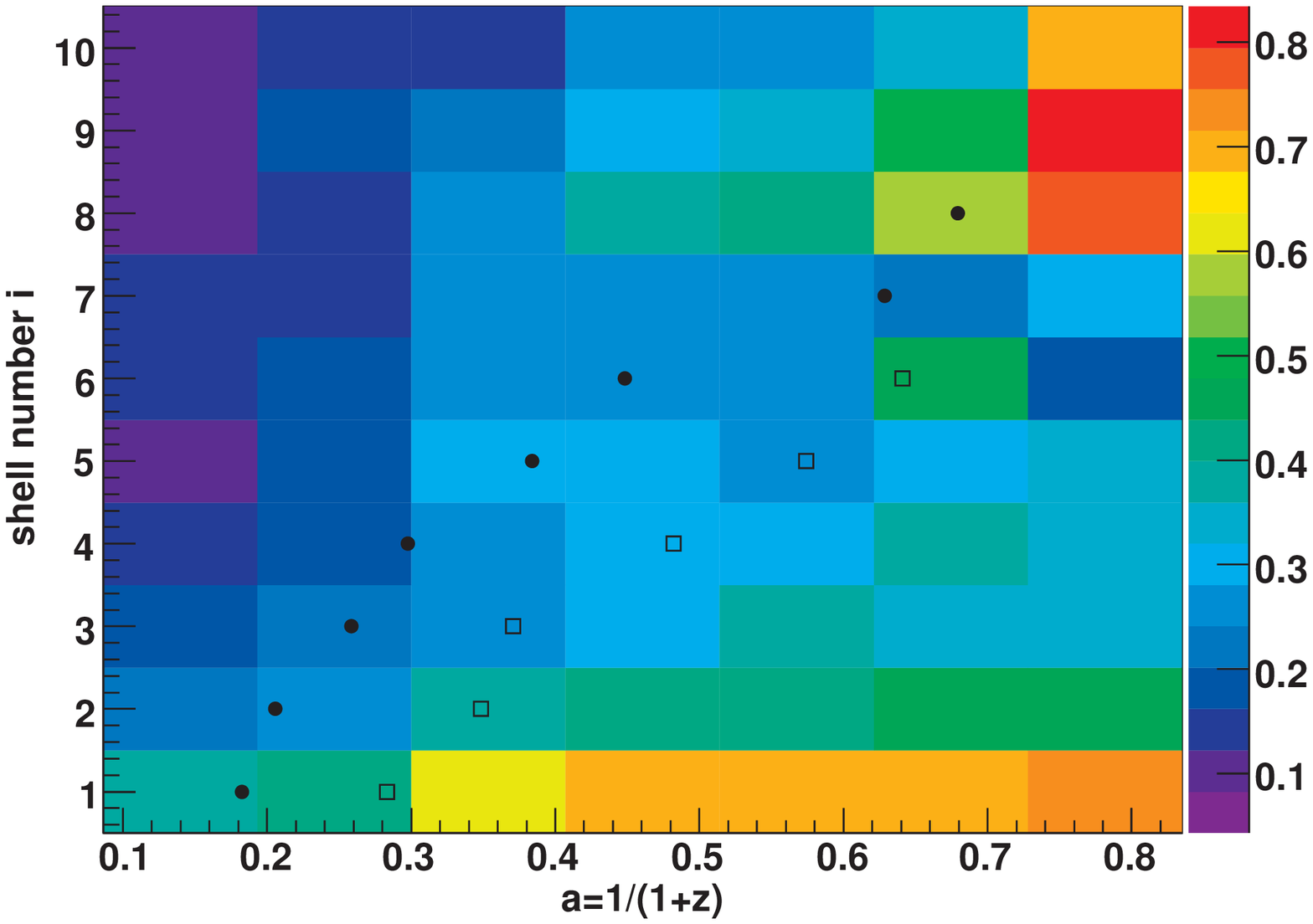,width=0.8\textwidth}}
\vspace{0.0cm}
\caption{Fraction of material belonging to shell $i$ at epoch
$a$ that remains in the same shell today. Shells are same as in
Fig. \ref{LagRadii}, numbered from one (inner) to ten (outer).
{\it Solid circles}: time of maximum expansion. {\it Open squares:}
stabilization epoch. Mass mixing generally decreases with time and towards
the halo center.
}
\label{overlap}
\end{figure}

Following the spherical top-hat model, the common procedure used to describe the assembly of
a dark matter halo is to define at each epoch a virial radius $\rvir$ (or, equivalently,
$\rtwo$), which depends
on the cosmic background density at the time. As the latter decreases with the Hubble 
expansion, formal virial radii and masses grow with cosmic time even for stationary halos. Studying
the transformation of halo properties within $\rvir$ (or some fraction of it) mixes
real physical change with apparent evolutionary effects caused by the growing radial
window, and makes it hard to disentangle between the two.
Figure \ref{LagRadii} shows the formation of Via Lactea where radial shells enclosing a 
fixed mass, $r_M$, have been used instead. Unlike $\rvir$, $r_M$ stops 
growing as soon as the mass distribution 
of the host halo becomes stationary on the corresponding scale. Note that 
{\it mass and substructure 
are constantly exchanged between these shells}, as $r_M$ is
not a Lagrangian radius enclosing the same material at all times, just the same
amount of it. The fraction of material belonging to a given shell in the past
that still remains within the same shell today is shown in Figure \ref{overlap}. 
The mixing is larger before stabilization, presumably because of shell crossing 
during collapse, and smaller near the center, where most of the mass is in a 
dynamically cold, concentrated old component (Diemand \etal 2005b).
Outer shells number 9 and 10, for example, retain today less than 25\% of the 
particles that originally belonged to them at $a<0.4$.  
Note that the collapse times also appear to differ from the expectations of 
spherical top-hat. Shell number five, for example, encloses a mean density of
about $100\,\rho_c^0$ today, a virial mass of
$1.5\times10^{12}\,\msun$ and should have virialized just now
according to spherical top-hat. It did so instead much earlier, at
$a=0.6$. Even the next larger shell with $1.8\times10^{12} \msun$
stabilized before $a=0.8$. It appears that spherical top-hat provides only a crude
approximation to the virialized regions of simulated galaxy halos.

\begin{figure*}
\centerline{\epsfig{figure=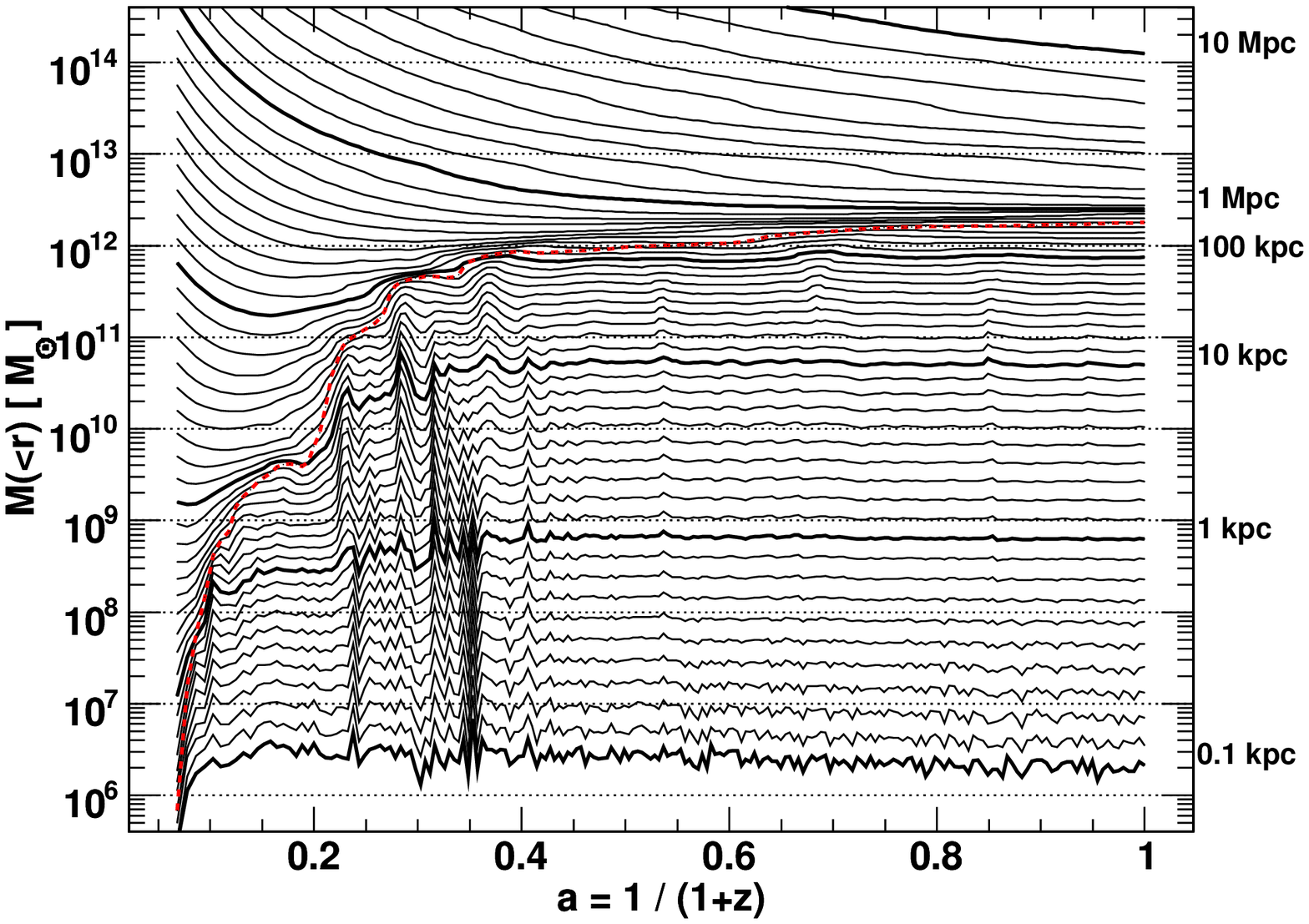,width=0.8\textwidth}}
\caption{Mass accretion history of Via Lactea. Masses within
spheres of fixed physical radii centered on the main progenitor are plotted
against the cosmological expansion factor $a$. The thick solid lines correspond
to spheres with radii given by the labels on the left. The thin solid lines
correspond to nine spheres of intermediate radii that are 1.3, 1.6, 2.0, 2.5, 3.2, 4.0,
5.0, 6.3 and 7.9 times larger than the next smaller labeled radius. {\it Dashed line:}
$M_{200}$. The halo is
assembled during a phase of active merging before $a\simeq0.37$ ($z\simeq1.7$) and
its net mass content remains practically stationary at later times.}
\label{massaccr_norvcm}
\end{figure*}

To understand the mass accretion history of the Via Lactea halo
it is useful to analyze the evolution of mass within fixed physical radii.
Figure \ref{massaccr_norvcm} shows that the mass within all radii from
the resolution limit of $\simeq $1 kpc up to 100 kpc grows during a series
of major mergers before $a=0.4$.  After this phase of active merging and
growth by accretion the halo mass distribution remains almost perfectly stationary at all
radii. Only the outer regions ($\sim $ 400 kpc) experience a small 
amount of net mass accretion after the last major merger. The mass within 400
kpc increases only mildly, by a factor of 1.2 from $z=1$ to the present.
During the same time the mass within radii of 100 kpc and smaller, the peak circular 
velocity and the radius where this is reached, all remain constant to within 10\%.
The fact that mass definitions inspired by spherical top-hat
fail to accurately describe the real assembly of galaxy halos is clearly seen in
Figure \ref{massaccr_norvcm}, where $M_{200}$ is shown to
increase at late times even when the halo physical mass remains the same.
This is just an artificial effect caused by the growing radial windows
$\rvir$ and $\rtwo$ as the background density decreases. For Via Lactea $M_{200}$
increases by a factor of 1.8 from $z=1$ to the present, while the real
physical mass within a 400 kpc sphere grows by only a factor of 1.2 during the
same time interval, and by an even smaller factor at smaller radii.

\subsection{Smallest  SUSY-CDM microhalos}

As already mentioned above, the key idea of the standard cosmological paradigm for the formation
of structure in the universe, that primordial density fluctuations
grow by gravitational instability driven by cold, collisionless dark
matter, is constantly being elaborated upon and explored in
detail through supercomputer simulations, and tested against a variety
of astrophysical observations. The leading candidate for dark matter
is the neutralino, a weakly interacting massive particle
predicted by the supersymmetry (SUSY) theory of particle
physics. While in a SUSY-CDM scenario the mass of bound dark matter
halos may span about twenty order of magnitudes, from the most
massive galaxy clusters down to Earth-mass clumps (Green et al. 2004), it 
is the smallest microhalos that
collapse first, and only these smallest scales are affected by the
nature of the relic dark matter candidate.
 
\begin{figure}
\vspace{0.4cm}
\centerline{\psfig{figure=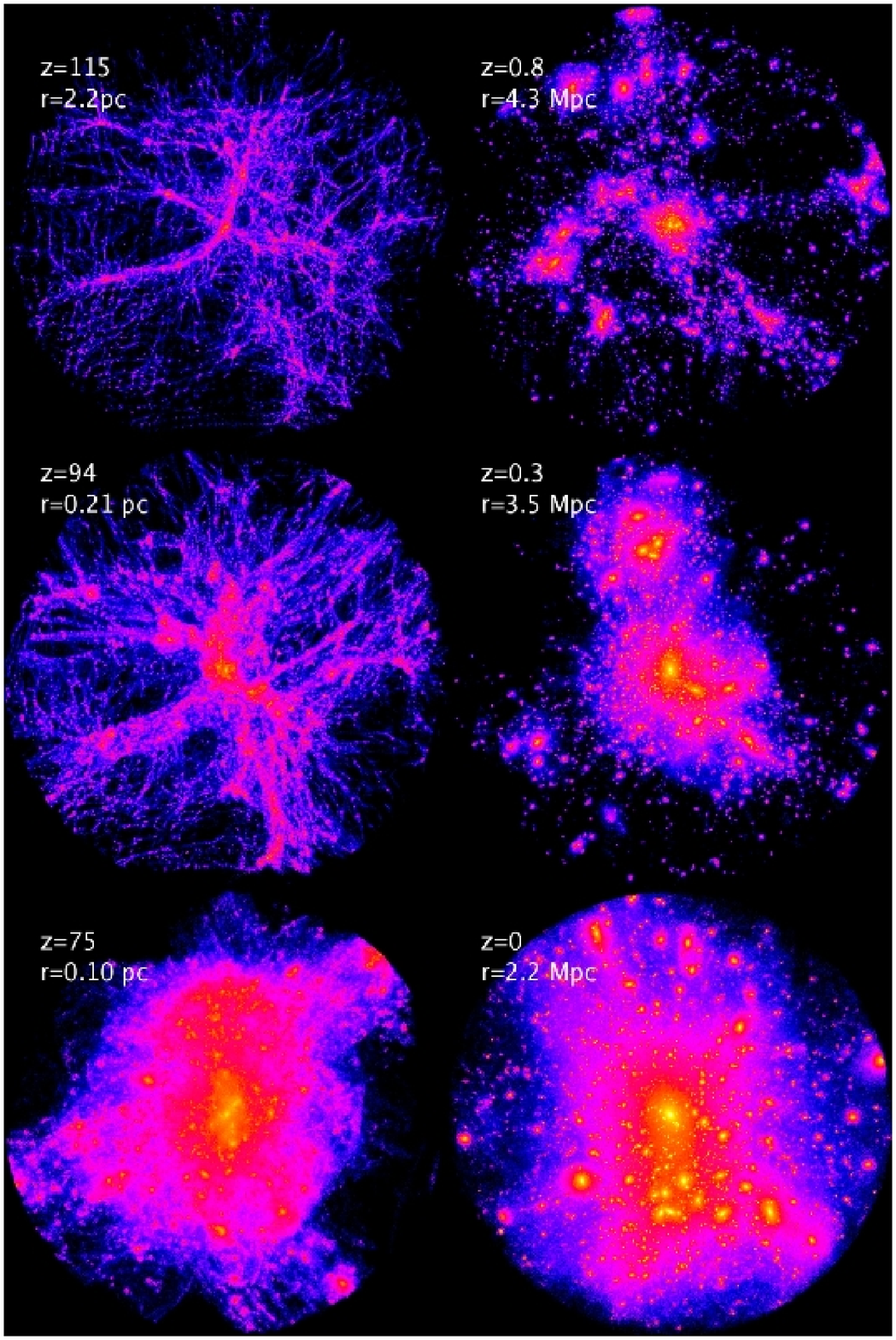,width=0.7\textwidth}}
\vspace{0.5cm}
\caption{Local dark matter density maps. The left panels illustrate the almost
simultaneous structure formation in the SUSY run at different epochs,
within a sphere of physical radius $r$ including a mass of $0.014\,\msun$. A galaxy 
cluster halo ({\it right panels})
forms in the standard hierarchical fashion: the dark matter distribution is shown within
a sphere of radius $r$ including a mass of $5.9\times 10^{14}\,\msun$.
The SUSY and cluster halos have concentration
parameter for a NFW profile of $c=3.7$ and $c=3.5$,
respectively. In each image the logarithmic color scale ranges from 10 to
$10^6\rho_c(z)$.
}
\label{sixpanel}
\end{figure}

Recent numerical simulations of the collapse of the earliest and
smallest gravitationally bound CDM clumps (Diemand et al. 2005a; Gao
et al. 2005) have shown that tiny virialized
microhalos form at redshifts above 50 with internal density profiles
that are quite similar to those of present-day galaxy clusters. At
these epochs a significant fraction of neutralinos has already been
assembled into non-linear Earth-mass overdensities. If this first
generation of dark objects were to survive gravitational disruption
during the early hierarchical merger and accretion process -- as well
as late tidal disruption from stellar encounters (Zhao \etal 2007) -- 
then over $10^{15}$ such clumps may populate the halo of the Milky Way. The
nearest microhalos may be among the brightest sources of $\gamma$-rays
from neutralino annihilation. As the annihilation rate increases
quadratically with the matter density, small-scale clumpiness may enhance
the total $\gamma$-ray flux from nearby extragalactic systems (like
M31), making them detectable by the forthcoming {\it GLAST} satellite
or the next-generation of air Cerenkov telescopes.

The possibility of observing the fingerprints of the smallest-scale
structure of CDM in direct and indirect dark matter searches hinges on
the ability of microhalos to survive the hierarchical clustering
process as substructure within the larger halos that form at later
times. In recent years high-resolution N-body simulations have enabled
the study of gravitationally-bound subhalos with $\msub/M
\gta 10^{-6}$ on galaxy (and galaxy cluster) scales
(e.g. Moore \etal 1999; Klypin \etal 1999; Stoehr \etal 2003). The main
differences between these subhalos -- the surviving cores of objects
which fell together during the hierarchical assembly of galaxy-size
systems -- and the tiny sub-microhalos discussed here is that on the
smallest CDM scale the effective index of the linear power spectrum of
mass density fluctuations is close to $-3$. In this regime typical
halo formation times depend only weakly on halo mass, the capture of
small clumps by larger ones is very rapid, and sub-microhalos may be
more easily disrupted. 

\begin{figure}
\vspace{0.4cm}
\begin{center}
\includegraphics[width=.49\textwidth]{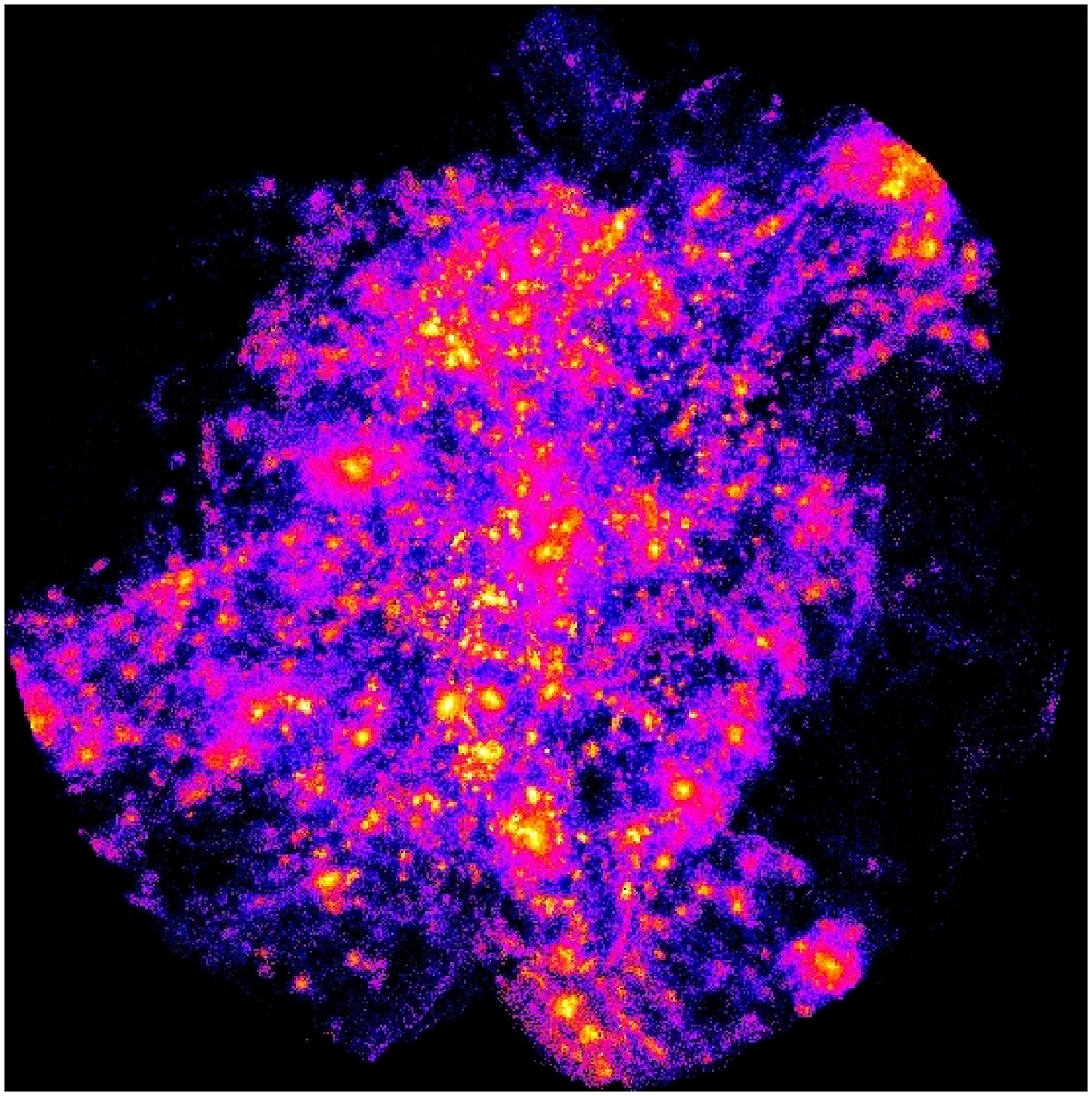}
\includegraphics[width=.49\textwidth]{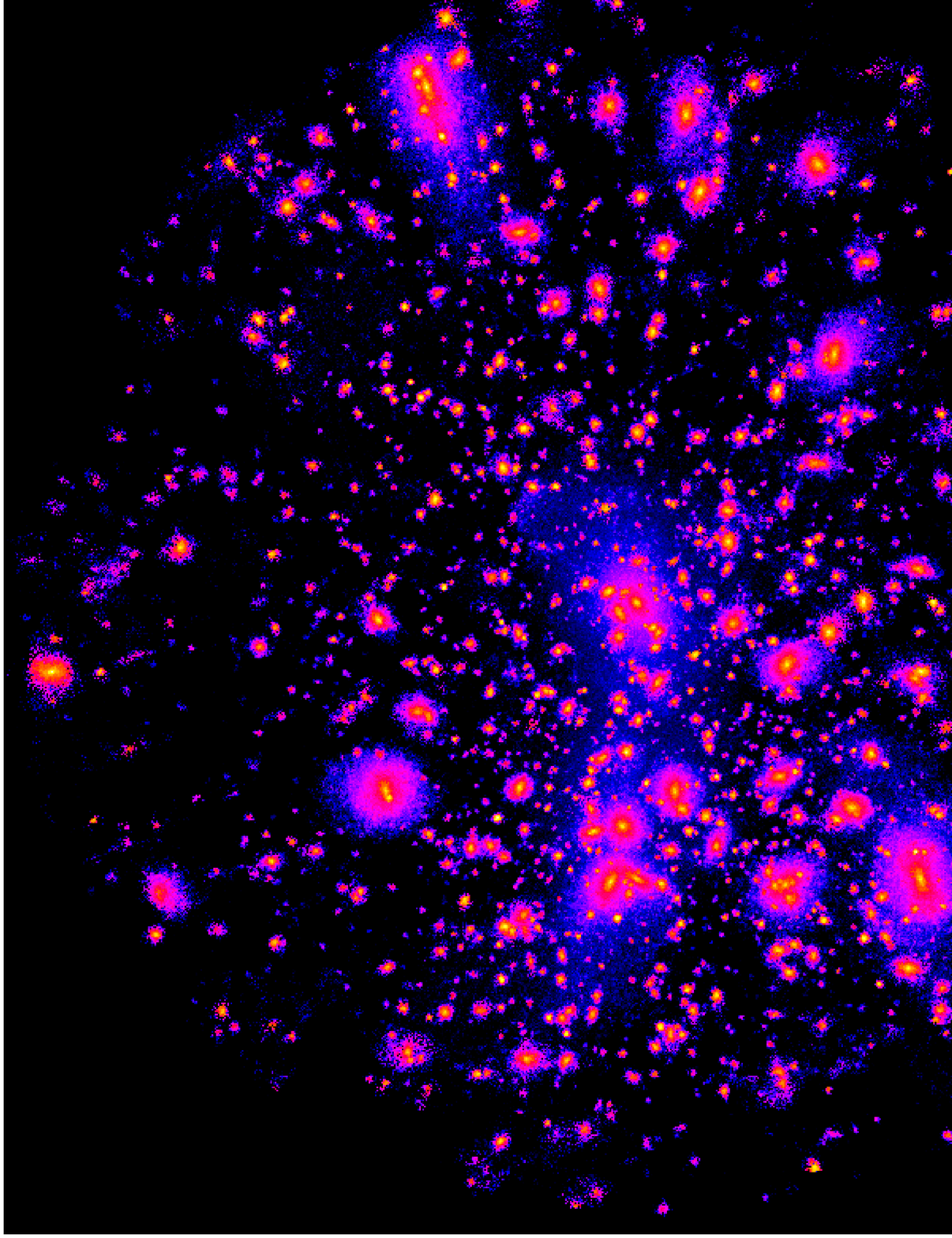}
\end{center}
\vspace{0.5cm}
\caption{Phase-space density ($\rho/\sigma^3$, where $\sigma$ is the one-dimensional 
velocity dispersion)  map for a $z=75$ SUSY halo ({\it left}) and a $z=0$ galaxy 
cluster ({\it right}). Note the different color scales: relative to the average
phase-space density, the logarithmic color
scale ranges from 10 to $10^{5}$ in the SUSY halo and from 10 to $10^7$ in the cluster
halo.}
\label{phasespace}
\end{figure}

In Diemand et al. (2006) we presented a large N-body simulation of
early substructure in a SUSY-CDM scenario characterized by an 
exponential cutoff in the power spectrum at $10^{-6}\,\msun$.
The simulation resolves a $0.014\,\msun$ parent ``SUSY'' halo at 
$z=75$ with 14 million particles. Compared to a $z=0$ galaxy cluster
substructure within the SUSY host is less evident both in phase-space 
and in physical space (see Figs. \ref{sixpanel} and \ref{phasespace}), 
and it is less resistant against tidal disruption. As the universe expands by 
a factor of 1.3, between 20 and 40 percent of well-resolved SUSY
substructure is destroyed, compared to only $\sim 1$ percent in the low-redshift
cluster. Nevertheless SUSY substructure is just as abundant as in $z=0$ galaxy
clusters, i.e. the normalized mass and circular velocity functions are very
similar.

\section{The Dawn of Galaxies}

\subsection{Uncertainties in the power spectrum} 

As mentioned in the Introduction, some shortcomings on galactic and
sub-galactic scales of the currently favored model of hierarchical
galaxy formation in a universe dominated by CDM have recently
appeared. The significance of these discrepancies is still debated,
and `gastrophysical' solutions involving feedback mechanisms may offer
a possible way out. Other models have attempted to solve the apparent
small-scale problems of CDM at a more fundamental level, i.e. by
reducing small-scale power.  Although the `standard' $\Lambda$CDM
model for structure formation assumes a scale-invariant initial power
spectrum of density fluctuations with index $n=1$, the
recent \textit{WMAP} data find strong evidence for a departure from
scale invariance, with a best fit value $n =
0.951_{-0.019}^{+0.015}$. Furthermore the 1-year \textit{WMAP}
results favored a slowly varying spectral index, $dn/d\ln
k=-0.031^{+0.016}_{-0.018}$, i.e. a model in which the spectral index
varies as a function of wavenumber $k$ (Spergel \etal 2003). In the
3-year \textit{WMAP} data such a `running spectral index' leads to
a marginal improvement in the fit. Models with either $n<1$ or
$dn/d\ln k<0$ predict a significantly lower amplitude of fluctuations
on small scales than standard $\Lambda$CDM. The suppression of
small-scale power has the advantage of reducing the amount of
substructure in galactic halos and makes small halos form later (when
the universe was less dense) hence less  concentrated (Zentner \& Bullock 
2002).

\begin{figure}
\vspace{0.4cm}
\centerline{\psfig{figure=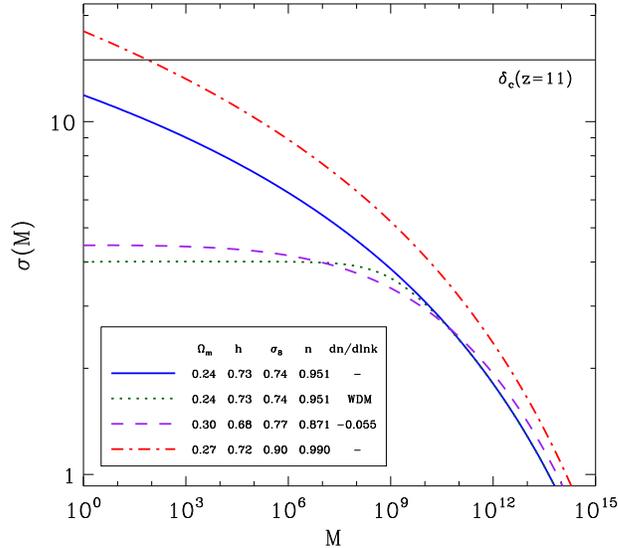,width=0.6\textwidth}}
\caption{The variance of the matter-density field vs. mass $M$, for several
different cosmologies, all based on \textit{WMAP}
results. \textit{Solid curve:} 3-year \textit{WMAP}-only
best-fit model. \textit{Dotted curve:} $\Lambda$WDM with a particle
mass $m_X=2$ keV, otherwise same as before. \textit{Dashed curve:}
3-year year \textit{WMAP} best-fit running spectral index
model. \textit{Dash-dotted curve:} 1-year \textit{WMAP}-only best-fit 
tilted model. Here $n$ refers to the
spectral index at $k=0.05\,\mbox{Mpc}^{-1}$. The horizontal line at
the top of the figure shows the value of the extrapolated critical collapse
overdensity $\delta_c(z)$ at the reionization redshift $z=11$.
\label{fig:sigmaM}
}
\end{figure}

Figure~\ref{fig:sigmaM} shows the linearly extrapolated (to $z=0$)
variance of the mass-density field
for a range of cosmological parameters. Note that
the new \textit{WMAP} results prefer a low value for $\sigma_8$, the
rms mass fluctuation in a 8 $h^{-1}\,$Mpc sphere. This is consistent
with a normalization by the $z=0$ X-ray cluster abundance (Reiprich \& B\"oringer 2002).
For comparison we have also included a model
with a higher normalization, $\sigma_8=0.9$, from the best-fit model to
1-year \textit{WMAP} data. As in the CDM paradigm structure formation proceeds 
bottom-up, it then follows that the loss of small-scale
power modifies structure formation most severely at the highest
redshifts, significantly reducing the number of self-gravitating
objects then. This, of course, will make it more difficult to reionize
the universe early enough.

\begin{figure}
\begin{center}
\includegraphics[width=.49\textwidth]{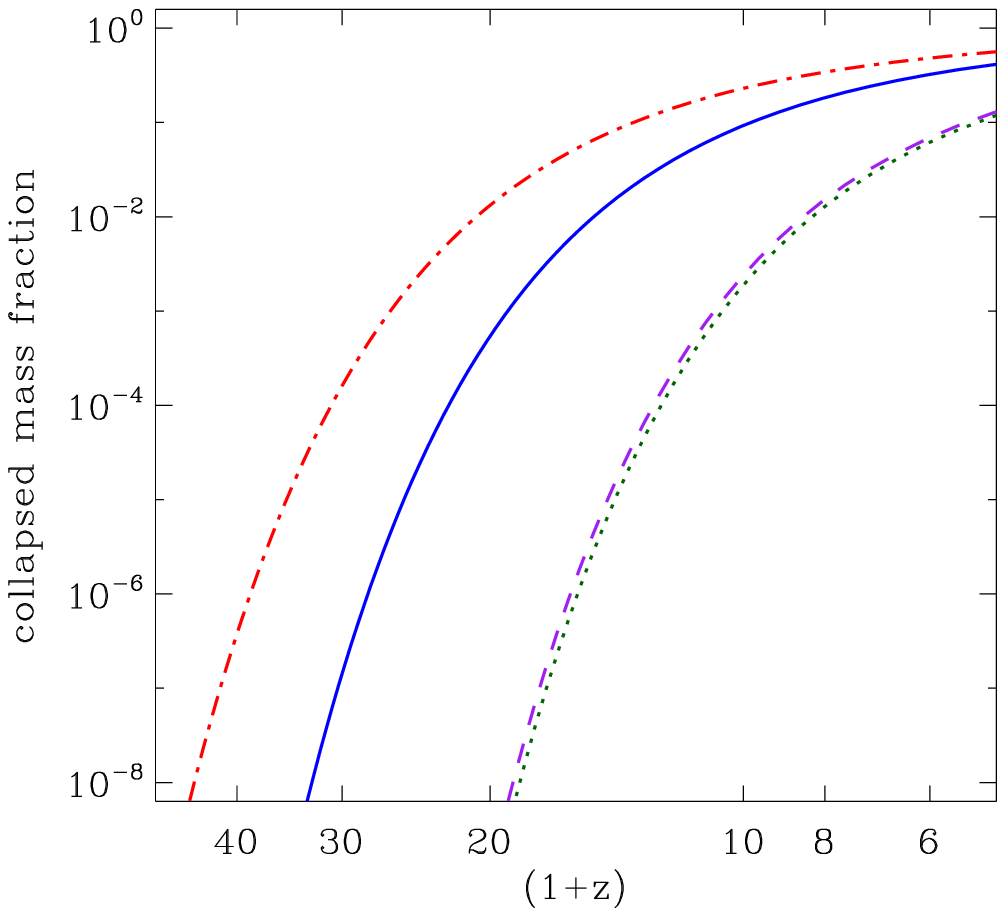}
\includegraphics[width=.49\textwidth]{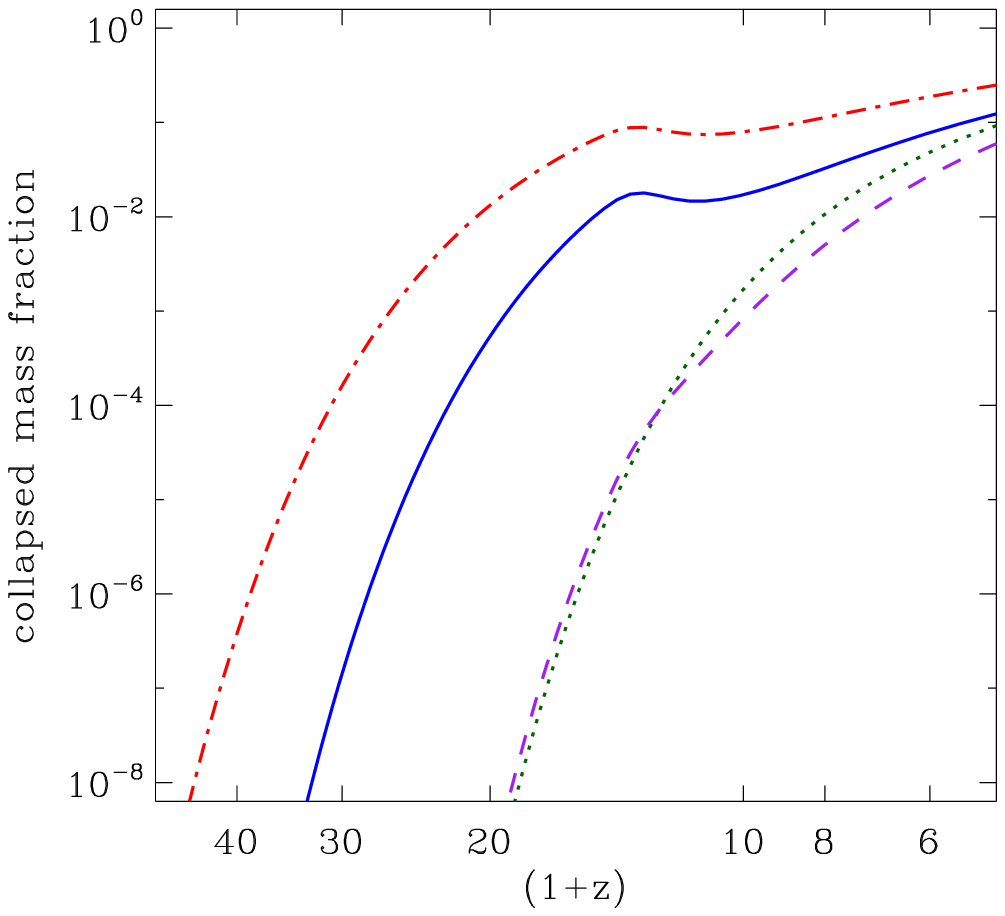}
\end{center}
\caption{Mass fraction in all
collapsed halos above the cosmological filtering (Jeans) mass as a function of
redshift, for different power spectra.  Curves are the same as in
Figure~\ref{fig:sigmaM}. \textit{Left panel:} filtering mass computed 
in the absence of reionization. \textit{Right panel}:
filtering mass computed assuming the universe is reionized and reheated to
$T_e=10^4$ K by ultraviolet radiation at $z\simeq 11$.
\label{fig:collapsed}
}
\end{figure}

It has been argued, for example, that one popular modification of the
CDM paradigm, warm dark matter (WDM), has so little structure at high
redshift that it is unable to explain the \textit{WMAP} observations
of an early epoch of reionization (Barkana \etal 2001). And
yet the \textit{WMAP} running-index model may suffer from a similar
problem.  A look at Figure~\ref{fig:sigmaM}
shows that $10^6\,$M$_\odot$ halos will collapse at $z=11$ from
$2.4\,\sigma$ fluctuations in a tilted $\Lambda$CDM model with
$n=0.951$ and $\sigma_8=0.74$ (best-fit 3-year \textit{WMAP}
model), but from much rarer 3.7 and 3.6 $\sigma$ fluctuations in the
WDM and running-index model, respectively. The problem is that
scenarios with increasingly rarer halos at early times require even
more extreme assumptions (i.e. higher star formation efficiencies and
UV photon production rates) in order to be able to reionize the
universe suitably early (e.g. Somerville \etal 2003; Wyithe \& Loeb 2003; 
Ciardi \etal 2003; Cen 2003).
Figure~\ref{fig:collapsed} depicts the mass fraction in all collapsed
halos with masses above the cosmological filtering mass for a case without
reionization and one with reionization occurring at $z\simeq 11$.  At
early epochs this quantity appears to vary by orders of magnitude in
different models.

\subsection{First baryonic objects} 

\begin{figure}
\centerline{\psfig{figure=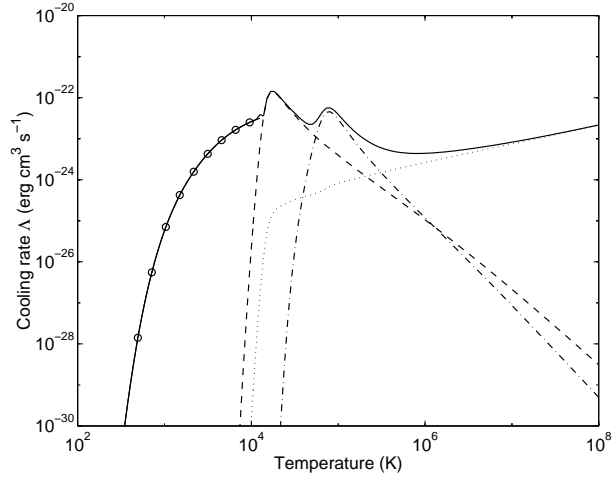,width=0.6\textwidth}}
\caption{Cooling rates for Bremsstrahlung ({\it dotted line}), H ({\it dashed line}) and 
He ({\it dash-dotted line}) line cooling, and \HH\ ({\it circles}) cooling. The $e^-$, \HII,
\HeII, and \HeIII\ abundances were computed assuming collisional equilibrium, and the 
\HH\ fractional abundance was fixed at $3\times 10^{-4}$, which is typical 
for early objects. At temperatures between 100 and 10,000 K, the \HH\ molecule is the
most effective coolant (From Fuller \& Couchman 2000).  
\label{crate}
}
\end{figure}

\begin{figure}
\centerline{\psfig{figure=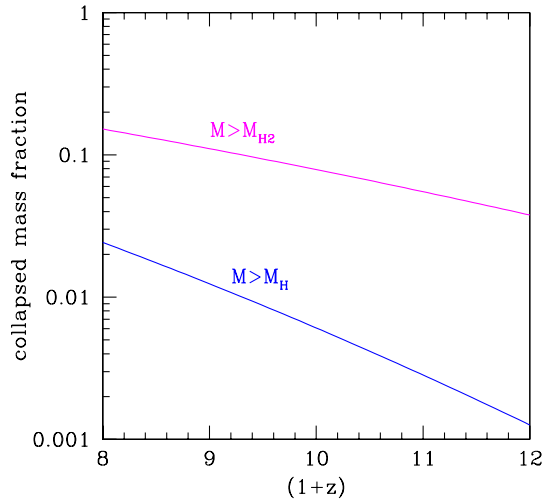,width=0.6\textwidth}}
\caption{{\it Solid lines:} Total mass fraction in all
collapsed dark matter
halos above the molecular cooling and the atomic cooling masses, $M_{\rm H_2}$
and $M_{\rm H}$,
as a function of redshift.
\label{figcool}
}
\end{figure}

The study of the non-linear regime for the baryons is far more complicated 
than that of the dark matter because of the need to take into
account pressure gradients and radiative processes. As a dark matter halo 
grows and virializes above the cosmological Jeans mass through merging and accretion, 
baryonic material will be shock heated to the effective virial temperature of 
the host and compressed to the same fractional overdensity as the dark matter. 
The subsequent behavior of gas in a dark matter halo depends on the efficiency with 
which it can cool. It is useful here to
identify two mass scales for the host halos: (1) a {\it molecular cooling mass} $M_{\rm H_2}$
above which gas can cool via roto-vibrational levels of H$_2$ and contract,
$M_{\rm H_2}\approx 10^5\, [(1+z)/10]^{-3/2}\,\msun$ (virial temperature
above $200\,$K); and (2) {\it an atomic cooling mass} $M_{\rm H}$ above which
gas can cool efficiently and fragment via excitation of hydrogen Ly$\alpha$,
$M_{\rm H} \approx 10^8\, [(1+z)/10]^{-3/2}\,\msun$ (virial temperature above
$10^4\,$K). Figure \ref{crate} shows the cooling mechanisms at various temperatures
for primordial gas, while Figure \ref{figcool} shows the fraction of the total mass in the
universe that is in collapsed dark matter halos with masses greater than
$M_{\rm H_2}$ and $M_{\rm H}$ at different epochs.

High-resolution hydrodynamics simulations of early structure formation
are a powerful tool to track in detail the thermal and ionization history 
of a clumpy IGM and guide studies of primordial star formation and reheating. 
Such simulations performed in the context of $\Lambda$CDM cosmologies have shown 
that the first stars (the so-called `Population III') in the universe formed out of
metal-free gas in dark matter minihalos of mass above a few $\times 10^5\,\msun$ 
(Abel \etal 2000; Fuller \& Couchman
2000; Yoshida \etal 2003; Reed \etal 2005) condensing from the rare
high-$\nu_c$ peaks of the primordial density fluctuation field at
$z>20$, and were likely very massive (e.g. Abel \etal 
2002; Bromm \etal 2002; see Bromm \& Larson 2004 and Ciardi \& Ferrara
2005 for recent reviews). In Kuhlen \& Madau (2005) we used a modified version of \textsc{enzo}, an
adaptive mesh refinement (AMR), grid-based hybrid (hydro$+$N-body)
code developed by Bryan \& Norman (see http://cosmos.ucsd.edu/enzo/)
to solve the cosmological hydrodynamics equations and study the cooling
and collapse of primordial gas in the first baryonic structures.
The simulation samples the dark matter density field in a 0.5 Mpc box with 
a mass resolution of $2000\,\msun$ to ensure
that halos above the cosmological Jeans mass are well
resolved at all redshifts $z<20$. The AMR technique allows us to home in, 
with progressively finer resolution, on the densest parts of the ``cosmic web".
During the evolution from $z=99$ to $z=15$, refined (child) grids are
introduced with twice the spatial resolution of the coarser (parent)
grid when a cell reaches a dark matter overdensity 
(baryonic overdensity) of 2.0 (4.0). Dense regions are allowed to 
dynamically refine to a maximum resolution of 30 pc (comoving).
The code evolves the non-equilibrium rate
equations for nine species (H, H$^+$, H$^-$, e, He, He$^+$, He$^{++}$,
\HH, and \HH$^+$) in primordial self-gravitating gas, including
radiative losses from atomic and molecular line cooling, 
and Compton cooling by the cosmic background radiation. 

\begin{figure}
\includegraphics[width=0.99\textwidth]{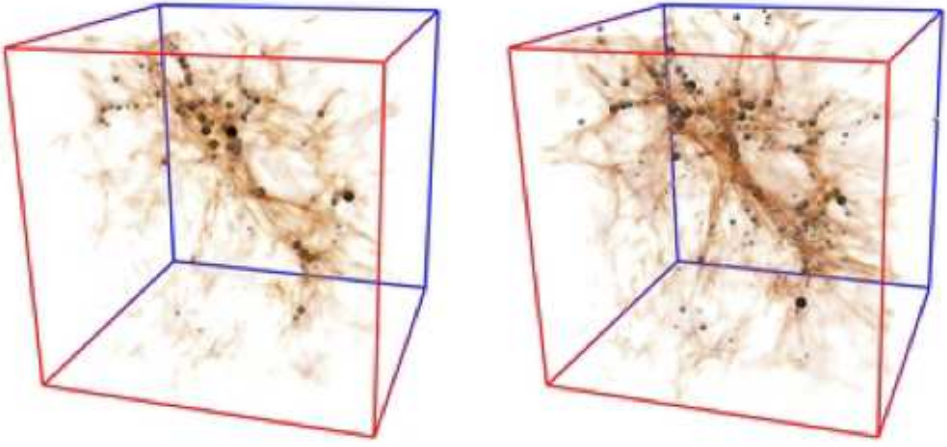}
\caption{A 3D volume rendering of the IGM in the inner 0.5 Mpc simulated box at
$z=21$ ({\it left panel}) and $z=15.5$ ({\it right panel}). Only gas with 
overdensity $4<\delta_b<10$ is shown:
the locations of dark matter minihalos are marked by spheres with
sizes proportional to halo mass. At these
epochs, the halo finder algorithm identifies 55 ($z=21$) and 262
($z=15.5$) bound clumps within the simulated volume. (From Kuhlen \& Madau 2005.)
}
\label{fig:3Dvol}
\end{figure}

\begin{figure}
\includegraphics[width=0.99\textwidth]{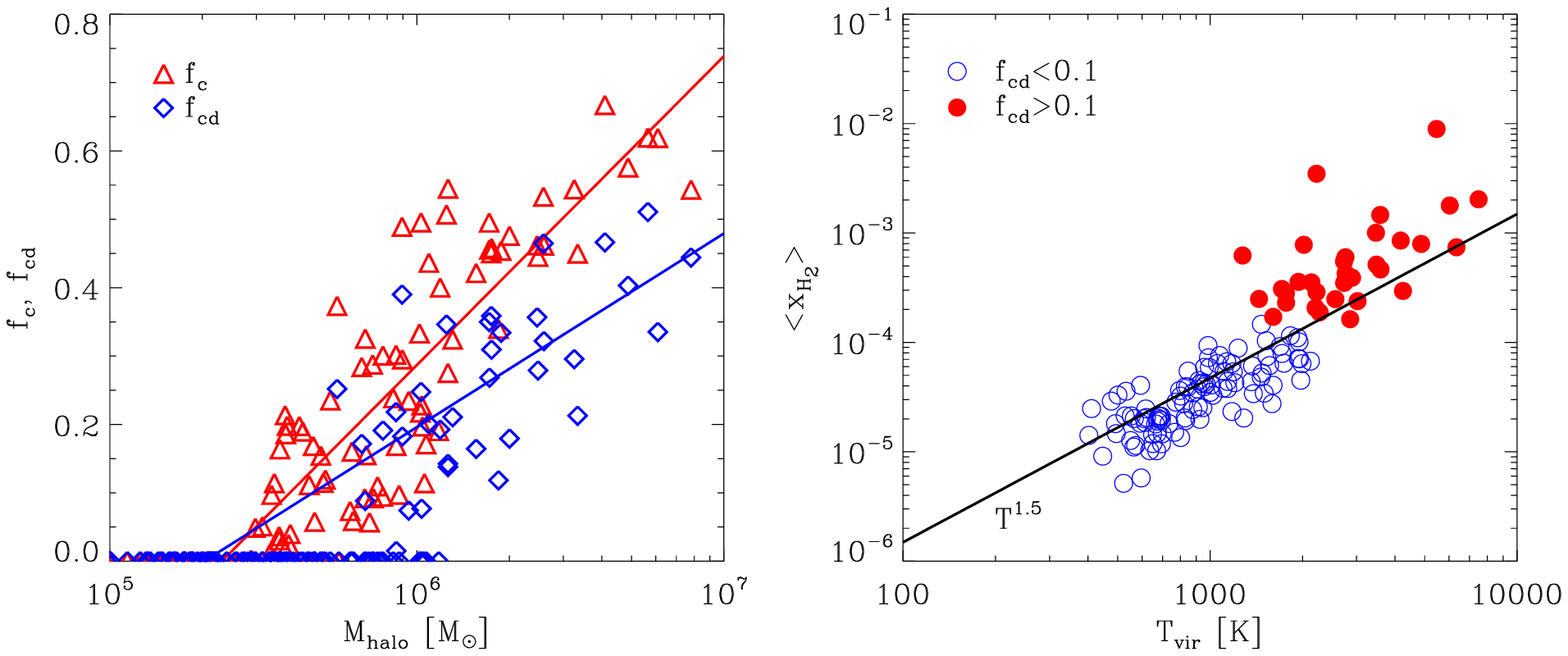}
\caption{
{\it Left:} Fraction of cold and cold$+$dense gas within the virial
radius of all halos indentified at $z=17.5$ with $\Tvir>400$ K, as a
function of halo mass. {\it Triangles:} $f_c$, fraction of halo gas
with $T<0.5\,\Tvir$ and overdensity $\delta_b>1000$ that cooled via roto-vibrational
transitions of \HH.  {\it Diamonds:} $f_{\rm cd}$, fraction of gas
with $T<0.5\,\Tvir$ and $\rho> 10^{19}\,\msun$ Mpc$^{-3}$ that is
available for star formation. The straight lines represent mean
regression analyses of $f_c$ and $f_{\rm cd}$ with the logarithm of
halo mass. {\it Right:} Mass-weighted mean \HH\ fraction as a function
of virial temperature for all halos at $z=17.5$ with $\Tvir>400\,$K
and $f_{\rm cd}<0.1$ ({\it empty circles}) or $f_{\rm cd}>0.1$ ({\it
filled circles}). The straight line marks the scaling of the
temperature-dependent asymptotic molecular fraction.
}
\label{fig:coldense_noBH}
\end{figure}

The clustered structure around the most massive peaks
of the density field is clearly seen in Figure \ref{fig:3Dvol}, a
3D volume rendering of the simulated volume at redshifts 21 and
15.5. The figure shows gas at $4<\delta_b<10$, with the locations of
dark matter minihalos marked by spheres colored and sized according
to their mass (the spheres are only markers, the actual shape of the
halos is typically non-spherical). Several interleaving filaments are
visible, at the intersections of which minihalos are typically
found. At $z=21$, 55 bound halos are indentified in the simulated volume:
by $z=17.5$ this number has grown to 149, and by $z=15.5$ to 262 halos. 
At this epoch, only four halos have reached the critical virial 
temperature for atomic cooling, $\Tvir=10^4\,$K.

The primordial fractional abundance of \HH\ in the low-density IGM is small,
$x_{\rm H_2}\equiv [{\rm H_2/H}]\simeq 2\times 10^{-6}$, as at $z>100$ \HH\ formation is
inhibited because the required intermediaries, either \HH$^+$ or
H$^-$, are destroyed by CMB photons. Most of the gas in the simulation
therefore cools by adiabatic expansion. Within collapsing minihalos,
however, gas densities and temperatures are large enough that \HH\
formation is catalyzed by H$^-$ ions through the associative
detachment reaction H+H$^-\longrightarrow$ \HH+$e^-$, and the molecular
fraction increases at the rate $dx_{\rm H_2}/dt\propto x_e n_{\rm HI}
\Tvir^{0.88}$, where $x_e$ is the number of electrons per hydrogen
atom. For $\Tvir \lta$ a few thousand kelvins the virialization shock
is not ionizing, the free electrons left over from recombination are
depleted in the denser regions, and the production of \HH\ stalls at a
temperature-dependent asymptotic molecular fraction $x_{\rm H_2}
\approx 10^{-8}\, \Tvir^{1.5}\,\ln(1+t/t_{\rm rec})$, where $t_{\rm
rec}$ is the hydrogen recombination time-scale (Tegmark \etal 1997). A
typical \HH\ fraction in excess of 200 times the primordial value is
therefore produced after the collapse of structures with virial
temperatures of order $10^3\,$K. This is large enough to efficiently
cool the gas and allow it to collapse within a Hubble time unless
significant heating occurs during this phase (Abel \etal 2000; Yoshida
\etal 2003).

Fig.~\ref{fig:coldense_noBH} (left panel) shows the fraction of cold
gas within the virial radius as a function of halo mass for all the
halos identified at redshift 17.5. Following Machacek \etal
(2001), we define $f_c$ as the fraction of gas with temperature
$<0.5\,\Tvir$ and density $>1000$ times the background (this is the
halo gas that is able to cool below the virial temperature because of
\HH), and $f_{\rm cd}$ as the fraction of gas with temperature
$<0.5\,\Tvir$ and (physical) density $>10^{19}\, \msun$ Mpc$^{-3}$
(this is the self-gravitating gas available for star formation). As in
Machacek \etal (2001), we find that both $f_c$ and $f_{\rm cd}$ are
correlated with halo mass. The threshold for significant baryonic 
condensation (non-zero $f_{\rm cd}$) is approximately $5\times 10^5\,\msun$ 
at these redshifts (Haiman \etal 1996).  Also depicted in
Fig.~\ref{fig:coldense_noBH} (right panel) is the mass-weighted mean
molecular fraction of all halos with $\Tvir>400\,$K. Filled circles
represent halos with $f_{\rm cd}>0.1$, while open circles represent
the others. The straight line marks the scaling of the asymptotic
molecular fraction in the electron-depletion transition regime. 
The maximum gas density reached at redshift 15 in the
most refined region of our simulation is $4\times 10^5\,$cm$^{-3}$
(corresponding to an overdensity of $3\times 10^8$): within this cold
pocket the excited states of \HH\ are in LTE and the cooling time is
nearly independent of density.

\subsection{21cm signatures of the neutral IGM}

It has long been known that neutral hydrogen in the diffuse IGM
and in gravitationally collapsed structures may be directly detectable
at frequencies corresponding to the redshifted 21cm line of hydrogen (Field 1959; Sunyaev \&
Zel'dovich 1975; Hogan \& Rees 1979, see Furlanetto \etal 2006 for a recent review). 
The emission or absorption of 21cm photons from neutral gas is governed by the spin 
temperature $T_S$, defined as
\be
n_1/n_0=3\exp(-T_*/T_S).
\ee
Here $n_0$ and $n_1$ are the number densities of hydrogen atoms in the singlet and
triplet $n=1$ hyperfine levels, and $T_*=0.068\,$K is the temperature corresponding
to the energy difference between the levels. To produce an absorption or emission
signature against the CMB, the spin temperature must
differ from the temperature of the CMB, $T=2.73\,(1+z)\,$K. At $30\lta z\lta 
200$, prior to the appearance of non-linear baryonic objects, the IGM cools adiabatically
faster than the CMB, spin-exchange collisions between hydrogen atoms couple $T_S$ to
the kinetic temperature $T_e$ of the cold gas, and cosmic hydrogen can be observed
in absorption (Scott \& Rees 1990; Loeb \& Zaldarriaga 2004).
At lower redshifts, the Hubble expansion rarefies the gas and makes collisions
inefficient: the spin states go into equilibrium with the radiation, and
as $T_S$ approaches $T$ the 21cm signal diminishes. It is the
first luminous sources that make uncollapsed gas in the universe
shine again in 21cm, by mixing the spin states either via \Lya scattering or
via enhanced free electron-atom collisions (e.g. Madau \etal 1997; Tozzi \etal 2000; 
Nusser 2005).

While the atomic physics of the 21cm transition is well understood in the cosmological
context, exact calculations of the radio signal expected during the era
between the collapse of the first baryonic structures and the epoch of complete
reionization have been difficult to obtain, as this depends on the spin temperature, 
gas overdensity, hydrogen neutral fraction, and line-of-sight peculiar velocities. 
When $T_S=T_e$, the visibility of the IGM at 21cm revolves around the quantity
$(T_e-T)/T_e$. If $T_e<T$, the IGM will appear in absorption
against the CMB; in the opposite case it will appear in emission.
To determine the kinetic temperature of the IGM
during the formation of the first sources, one needs a careful treatment of the
relevant heating mechanisms such as photoionization and shock heating.
In addition to the signal produced by the cosmic web, minihalos with virial
temperatures of a few thousand kelvins form in abundance at high redshift, and
are sufficiently hot and dense to emit collisionally-excited 21 cm
radiation (Iliev \etal 2002).

The \HI\ spin temperature is a weighted mean between $T_e$ and $T$,
\be
T_S=\frac{T_*+T+yT_e}{1+y},  \label{eq:Tspin}
\ee
where coupling efficiency $y$ is the sum of three terms,
\be
y={T_*\over A T_e}\,(C_{\rm H}+C_e+C_p).
\ee
Here $A=2.85\times 10^{-15}\,$s$^{-1}$ is the spontaneous emission rate and
$C_{\rm H}$, $C_e$, and $C_p$ are the de-excitation rates of the
triplet due to collisions with neutral atoms, electrons, and protons.
A fourth term must be added in the presence of ambient \Lya radiation,
as intermediate transitions to the $2p$ level can mix the spin states and
couple $T_S$ to $T_e$, the ``Wouthuysen-Field'' effect (Wouthuysen 1952; Field 1958;
Hirata 2006). In the absence of \Lya photons, to unlock the spin temperature of a neutral medium
with (say) $T_e=500\,$K from the CMB requires a collision rate $C_{\rm H}>AT/T_*$, 
corresponding to a baryon overdensity $\delta_b>5\,[(1+z)/20]^{-2}$. Not only dense gas
within virialized minihalos but also intergalactic filamentary structure heated
by adiabatic compression or shock heating may then be observable in 21cm emission.
A population of X-ray sources (``miniquasars'') turning on at early stages (e.g. 
Madau \etal 2004; Ricotti \etal 2005) may make even the low-density IGM visible in 
21cm emission as structures develop in the pre-reionization era.

\begin{figure}
\vspace{0.2cm}
\centerline{\epsfig{figure=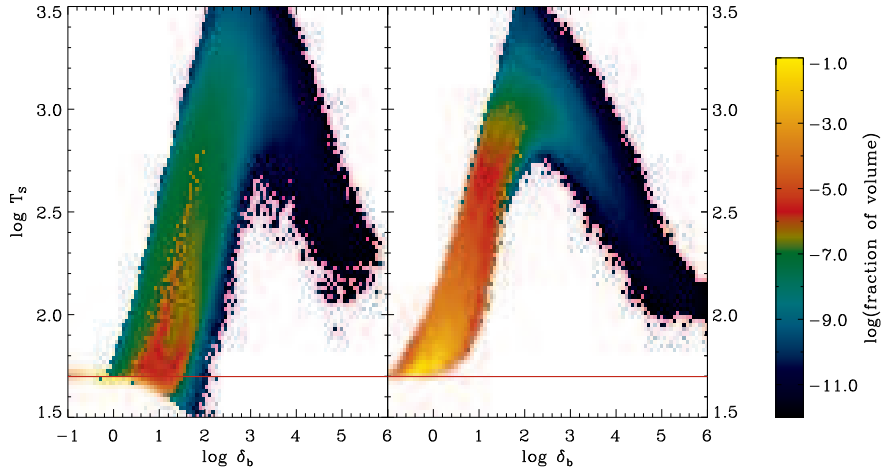,width=0.9\textwidth}}
\caption{\footnotesize Two-dimensional distribution of spin temperature versus
baryonic overdensity at $z=17.5$. The color coding indicates the
fraction of the simulated volume at a given ($\delta_b,T_S)$. {\it
Left:} NS run. The volume and mass-averaged spin temperatures are
48.6 and 67.5 K, respectively.  {\it Right:} MQ run. Only gas with
neutral fraction $>90\%$ is shown in the figure. The volume and
mass-averaged spin temperatures are 82.6 and 138.0 K,
respectively. The red horizontal line marks the temperature of
the CMB at that redshift. \label{fig1}}
\end{figure}

In Kuhlen \etal (2006) we used the same hydrodynamical simulations of early structure
formation in a $\Lambda$CDM universe discussed in \S~1.4.2
to investigate the spin temperature and 21cm brightness of the diffuse IGM
prior to the epoch of cosmic reionization, at $10<z<20$. 
The two-dimensional distribution of gas overdensity and spin
temperature at $z=17.5$ is shown in Figure~\ref{fig1} for a simulation
with no radiation sources (``NS") and one (``MQ") in which a
miniquasar powered by a $150\,\msun$ black hole turns on at redshift
21 within a host halo of mass $2\times 10^6\,\msun$. The miniquasar
shines at the Eddington rate and emits X-ray radiation with a
power-law energy spectrum $\propto E^{-1}$ in the range $0.2-10$ keV.
The color coding in this phase diagram indicates the fraction of the
simulated volume at a given ($\delta_b,T_S)$. In both runs we have
assumed no \Lya mixing, so that {\it the visibility of hydrogen at
21cm is entirely determined by collisions.} Only gas with neutral
fraction $>90\%$ is shown in the figure. The low-density IGM in the
NS run lies on the yellow $T_S=T=50.4\,$K line: this is
gas cooled by the Hubble expansion to $T_e\ll T$ that cannot
unlock its spin states from the CMB and therefore remains invisible.
At overdensities between a few and $\sim 200$, H-H collisions become
efficient and adiabatic compression and shocks from structure
formation heats up the medium well above the radiation
temperature. The coupling coefficient at this epoch is $y\sim \delta_b
T_e^{-0.64}$: gas in this regime has $T<T_S\sim yT_e<T_e$
and appears in {\it emission} against the CMB (red and green
swath). Some residual hydrogen with overdensity up to a few
tens, however, is still colder than the CMB, and is detectable in
{\it absorption}. At higher densities, $y\gg 1$ and $T_S \rightarrow
T_e$: the blue cooling branch follows the evolutionary tracks in the
kinetic temperature-density plane of gas shock heated to virial
values, $T_e= 2000-10^4\,$K, which is subsequently cooling down to
$\sim 100\,$K because of \HH\ line emission. 

The effect of the miniquasar on the spin temperature is clearly
seen in the right panel of Figure~\ref{fig1}. X-ray radiation drives the
volume-averaged temperature and electron fraction ($x_e$) within the simulation box
from $(8\,{\rm K}, 1.4\times 10^{-4})$ to $(2800\,{\rm K}, 0.03)$, therefore
producing a warm, weakly ionized medium (Kuhlen \& Madau 2005). The H-H collision term for spin
exchange in the low-density IGM increases on the average by a factor $350^{0.36}\sim
8$, while the e-H collision term grows to $C_e\sim 0.5 C_{\rm H}$. Gas with
$(\delta_b, T_e, x_e)=(1, 2800\,{\rm K}, 0.03)$ has coupling efficiency $y=0.008$,
spin temperature $T_S=73\,{\rm K}>T$,
and can now be detected in emission against the CMB. Within 150 comoving kpc
from the source, the volume-averaged electron fraction rises above 10\%, and e-H
collisions dominates the coupling.

\begin{figure*}
\vspace{0.cm}
\begin{center}
\includegraphics[width=0.95\textwidth]{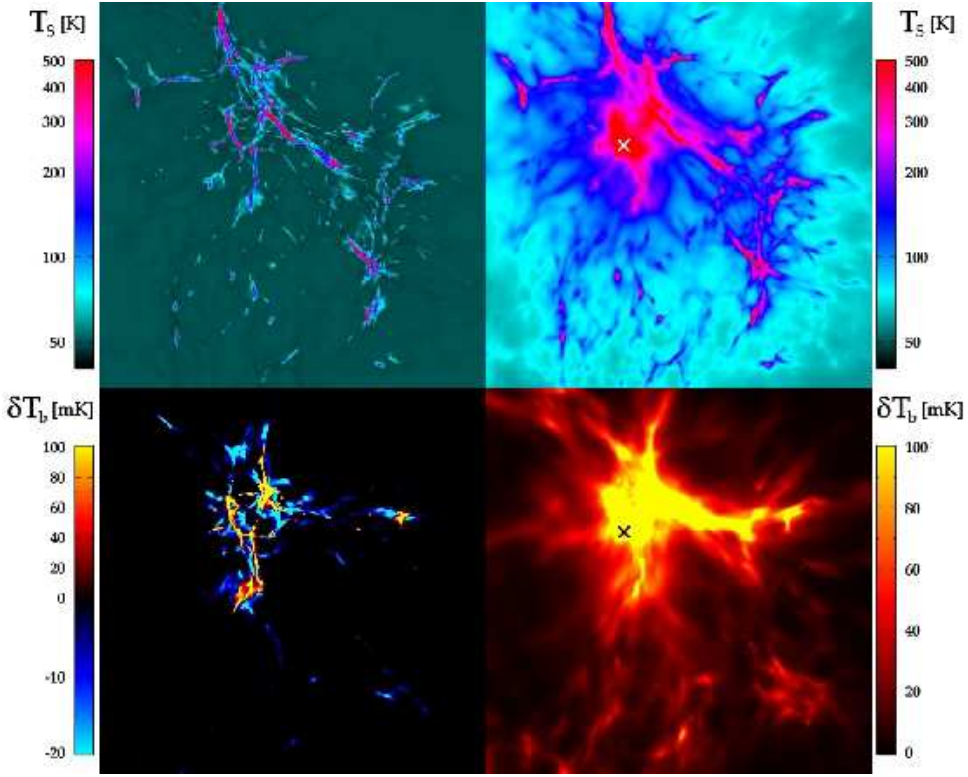}
\caption{\footnotesize Projected (mass-weighted) spin temperature
({\it upper panels}, logarithmic scale) and 21cm differential
brightness temperature ({\it lower panels}, composite linear scale) in a 0.5
Mpc simulation box for runs ``NS'' ({\it left}) and
``MQ'' ({\it right}) at $z=17.5$. The location of the miniquasar is
indicated by crosses in the right panels.} \label{fig2}
\end{center}
\end{figure*}
             
A beam of 21cm radiation passing through a neutral hydrogen patch having
optical depth $\tau$ and spin temperature $T_S$ causes absorption
and induces emission. In the comoving frame of the patch, the radiative transfer
equation yields for the brightness temperature through the region: $T_b=Te^{-\tau}+
T_S(1-e^{-\tau})$. We have used our numerical simulations to
perform 21cm radiation transport calculations, including the effect of peculiar
velocities and local changes in spin temperature, gas density, and neutral hydrogen
fraction. The resulting 21cm radio signal is shown in
Figure~\ref{fig2} (lower panels), together with an image of the {\it
projected} hydrogen spin temperature (upper panels): the latter
highlights the abundance of structure within our simulation box on
scales up to hundreds of kpc. Due to Hubble and peculiar velocity
shifts not all of this structure contributes to the $\delta T_b$
map. In the NS simulation, coherent features in the IGM can be
discerned in emission ($T_S>T$, $\delta T_b>0$): this
filamentary shock-heated structure is typically surrounded by mildly
overdense gas that is still colder than the CMB and appears in
absorption ($T_S<T$, $\delta T_b<0$). The covering factor of
material with $\delta T_b\le -10\,$ mK is 1.7\%, comparable to that of
material with $\delta T_b\ge +10\,$ mK: only about 1\% of the pixels
are brighter than +40 mK.  While low-density gas (black color in the
left-lower panel) remains invisible against the CMB in the NS run,
{\it the entire box glows in 21cm after being irradiated by X-rays.}
The fraction of sky emitting with $\delta T_b>(+10,+20,+30,+50,+100)$
mK is now (0.57,0.31,0.19,0.1,0.035).

The above calculations show that, even in the absence of
external heating sources, spin exchange by H-H collisions can make
filamentary structures in the IGM (heated by adiabatic compression or
shock heating) observable in 21cm emission at redshifts $z\lta 20$.
Some cold gas with overdensities in the range 5-100 is still
detectable in absorption at a level of $\delta T_b\lta -10\,$mK, with
a signal that grows as $T_S^{-1}$ and covers a few percent of the
sky. X-ray radiation from miniquasars preheats the IGM to a few
thousand kelvins and increases the electron fraction: this boosts both
the H-H and the e-H collisional coupling between $T_S$ and $T_e$,
making even low-density gas visible in 21cm emission well before the
universe is significantly reionized. Any absorption signal has
disappeared, and as much as 30\% of the sky is now shining with
$\delta T_b\gta +20\,$mK.  As pointed out by Nusser (2005), the
enhanced e-H coupling makes the spin temperature very sensitive to the
free-electron fraction: the latter is also a tracer of the \HH\
molecular fraction in the IGM. 

\subsection{Concluding remarks}

Since hierarchical clustering theories provide a well-defined
framework in which the history of baryonic material can be tracked
through cosmic time, probing the reionization epoch may then help
constrain competing models for the formation of cosmic structures.
Quite apart from uncertainties in the primordial power spectrum on
small scales, however, it is the astrophysics of baryons that makes us
unable to predict when reionization actually occurred. Consider the
following illustrative example:
  
Hydrogen photo-ionization requires more than one photon above 13.6 eV
per hydrogen atom: of order $t/t_{\rm rec}\sim 10$ (where $t_{\rm rec}$ is the 
volume-averaged hydrogen recombination timescale)
extra photons appear to be needed to keep the gas in overdense regions
and filaments ionized against radiative recombinations (Gnedin 2000; Madau 
\etal 1999).  A `typical' stellar population
produces during its lifetime about 4000 Lyman continuum (ionizing)
photons per stellar proton.  A fraction $f\sim 0.25$\% of cosmic
baryons must then condense into stars to supply the requisite
ultraviolet flux. This estimate assumes a standard (Salpeter) initial
mass function (IMF), which determines the relative abundances of hot,
high mass stars versus cold, low mass ones.
The very first generation of stars (`Population III') must have
formed, however, out of unmagnetized metal-free gas: characteristics, these,
which may have led to a `top-heavy' IMF biased towards very massive stars
(i.e. stars a few hundred times more massive than the Sun),
quite different from the present-day Galactic case. Pop III stars
emit about $10^5$ Lyman continuum photons per stellar baryon (Bromm \etal 2001), 
approximately $25$ times more than a standard
stellar population. A corresponding smaller fraction of cosmic baryons
would have to collapse then into Pop III stars to reionize the universe, $f\sim
10^{-4}$.  There are of course further complications. Since, at zero
metallicity, mass loss through radiatively-driven stellar winds is
expected to be negligible, Pop III stars
may actually die losing only a small fraction of their mass. If they
retain their large mass until death, stars with masses $140 \lta m_*
\lta 260\,\msun$ will encounter the electron-positron pair
instability and disappear in a giant nuclear-powered
explosion (Fryer \etal 2001), leaving no compact remnants and polluting
the universe with the first heavy elements. In still heavier stars,
however, oxygen and silicon burning is unable to drive an explosion,
and complete collapse to a black hole will occur instead (Bond \etal 1984). 
Thin disk accretion onto a Schwarzschild black hole
releases about 50 MeV per baryon. The conversion of a trace amount of
the total baryonic mass into early black holes, $f\sim 3\times
10^{-6}$, would then suffice to at least partially ionize and  reheat the universe.

The above discussion should make it clear that, despite much recent
progress in our understanding of the formation of early cosmic
structure and the high-redshift universe, the astrophysics of first
light remains one of the missing links in galaxy formation and
evolution studies. We are left very uncertain about the whole era from
$10^8$ to $10^9$ yr -- the epoch of the first galaxies, stars,
supernovae, and massive black holes.  Some of the issues
reviewed in these lectures
are likely to remain a topic of lively controversy until the
launch of the \textit{James Webb Space Telescope} (\textit{JWST}),
ideally suited to image the earliest generation of stars in the
universe.  If the first massive black holes form in pregalactic
systems at very high redshifts, they will be incorporated through a
series of mergers into larger and larger halos, sink to the center
owing to dynamical friction, accrete a fraction of the gas in the
merger remnant to become supermassive, and form binary systems (Volonteri \etal 2003). 
Their coalescence would be signaled by the
emission of low-frequency gravitational waves detectable by the
planned \textit{Laser Interferometer Space Antenna} (\textit{LISA}).
An alternative way to probe the end of the dark ages and discriminate
between different reionization histories is through 21cm tomography. 
Prior to the epoch of full reionization, 21cm
spectral features will display angular structure as well as structure
in redshift space due to inhomogeneities in the gas density field,
hydrogen ionized fraction, and spin temperature. Radio maps will show
a patchwork (both in angle and in frequency) of emission signals from
\HI\ zones modulated by \HII\ regions where no signal is detectable
against the CMB (Ciardi \& Madau 2003). The search at 21cm for the epoch
of first light has become one of the main science drivers for the
next generation of radio arrays.

\begin{figure}
\centerline{\epsfig{figure=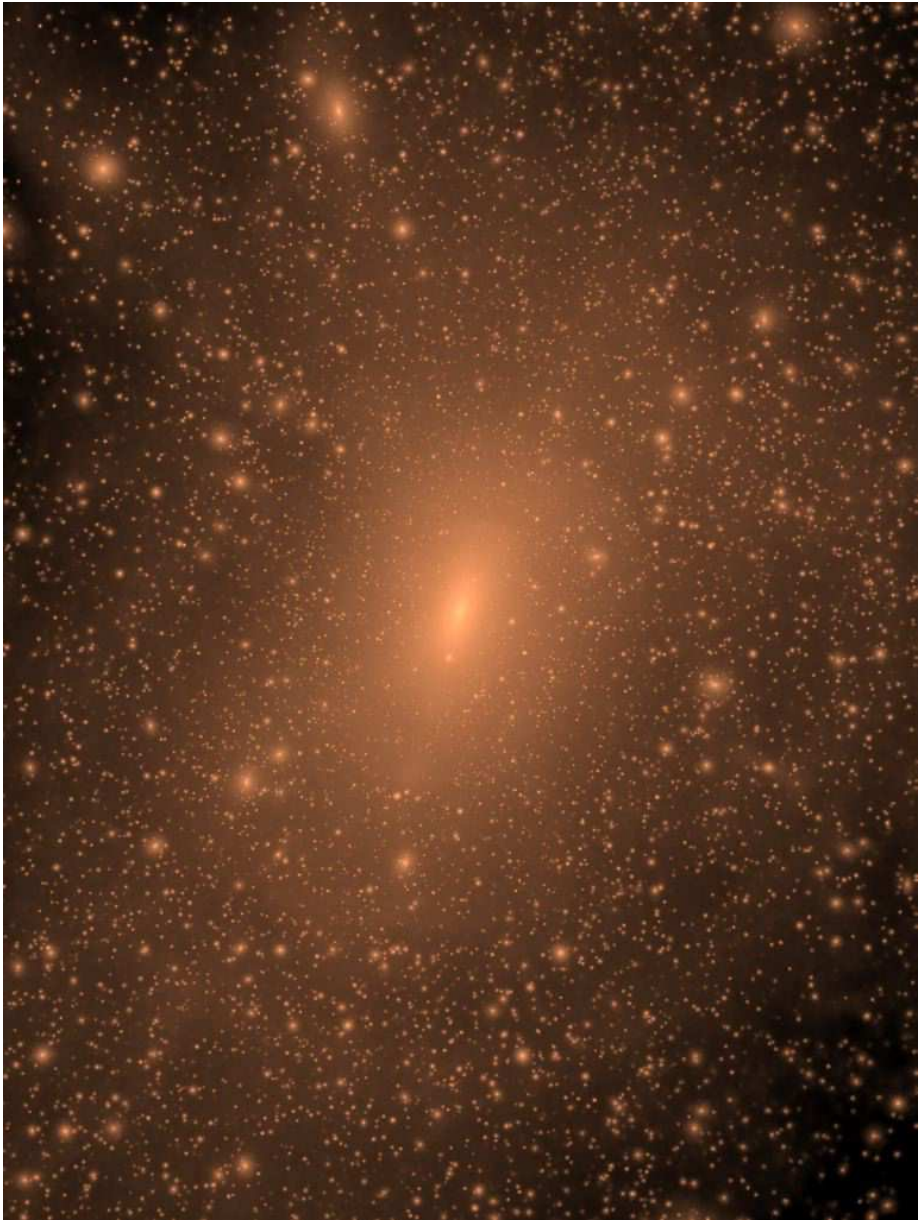,width=0.9\textwidth}}
\vspace{0.5cm}
\caption{Projected dark matter density-squared map of the Via Lactea halo
at the present epoch. The image covers
an area of 800 $\times$ 600 kpc, and the projection goes through a 600
kpc-deep cuboid containing a total of 110 million particles. The
logarithmic color scale covers 20 decades in density-square.  }
\label{VL}
\end{figure}

While many of the cosmological puzzles we have discussed can be tackled directly 
by studying distant objects, it has also become clear that many of today's ``observables'' 
within the Milky Way 
and nearby galaxies relate to events occurring at very high redshifts, during and soon 
after the epoch of reionization. In this sense, galaxies in the 
Local Group (``near-field cosmology'') can provide a crucial diagnostic link to the 
physical processes that govern structure formation and evolution in the early 
universe (``far-field cosmology''). It is now well established, for example, that 
the hierarchical mergers that form the halos surrounding galaxies are rather inefficient, leaving 
substantial amounts of stripped halo cores or ``subhalos'' orbiting
within these systems (see Fig. \ref{VL}). Small halos
collapse at high redshift when the universe is very dense, so their
central densities are correspondingly high. When these merge
into larger hosts, their high densities allow them to resist the
strong tidal forces that acts to destroy them. Gravitational
interactions appear to unbind most of the mass associated with 
the merged progenitors, but a significant fraction of these small 
halos survives as distinct substructure. 

\begin{figure}
\includegraphics[width=0.495\textwidth]{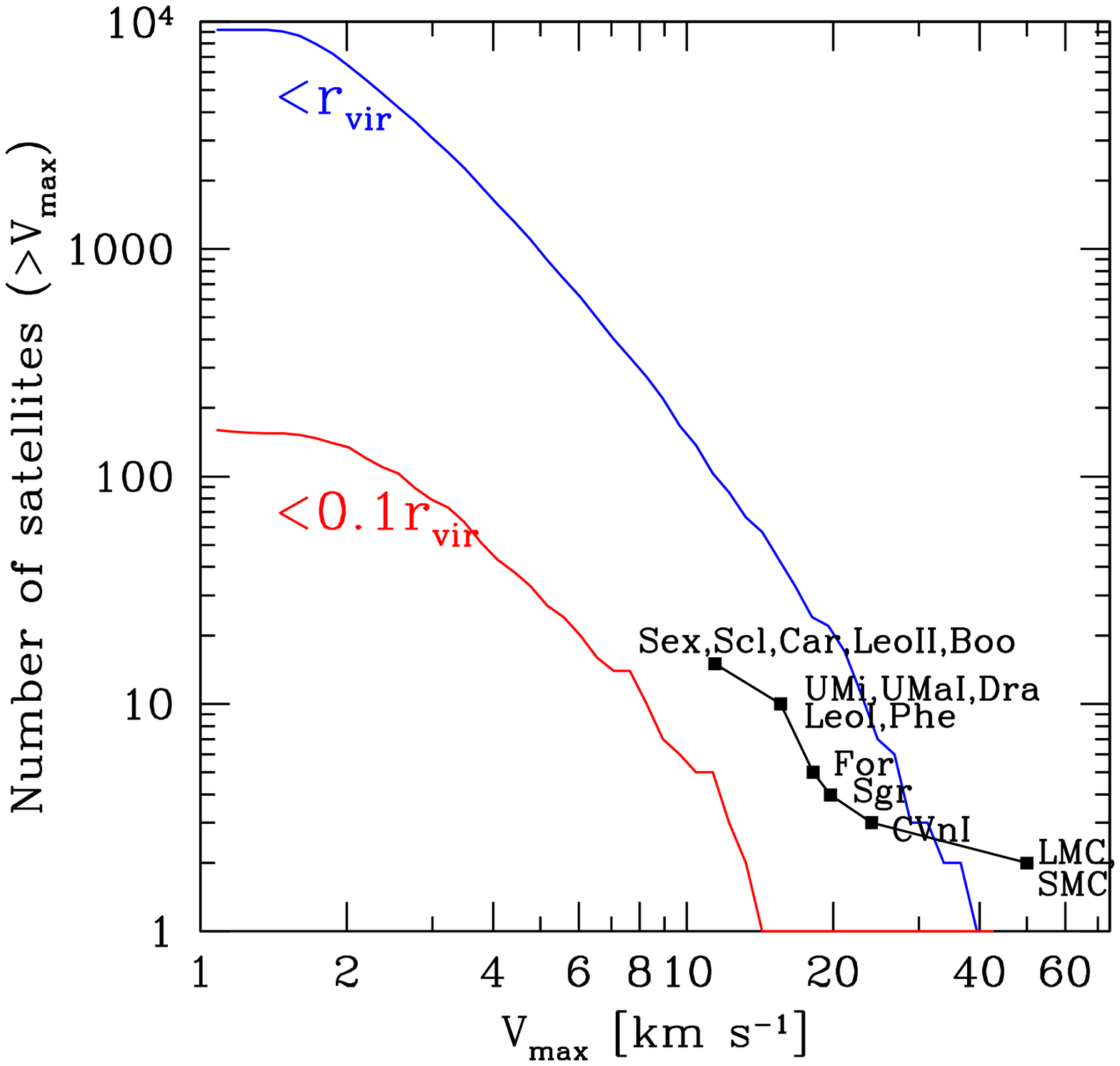}
\includegraphics[width=0.495\textwidth]{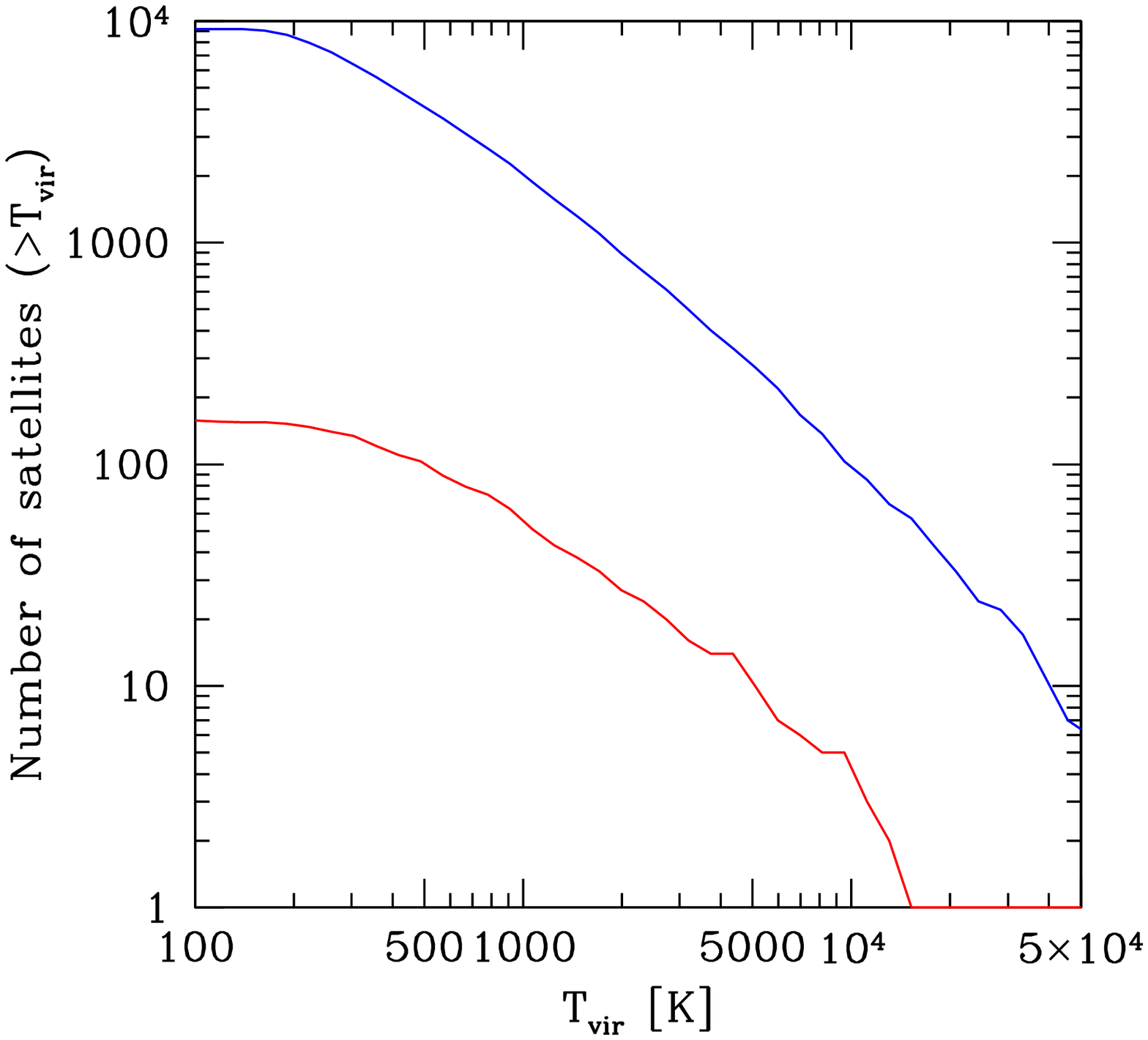}
\caption{{\it Left:} Cumulative peak circular velocity function for all subhalos within 
Via Lactea's $\rvir$ ({\it upper curve}) and for the subpopulation within 
the inner $0.1\,\rvir$ ({\it lower curve}). {\it Solid line with points:} observed 
number of dwarf galaxy satellites around the Milky Way. {\it Right:} Same plotted versus 
virial temperature $T_{\rm vir}$.}
\label{VLsat}
\end{figure}

The Via Lactea simulation has recently shown that, in the standard CDM paradigm, galaxy halos 
should be filled with tens of thousands subhalos that appear to have no optically luminous counterpart: 
this is more than an order of magnitude larger than found in previous 
simulations. Their cumulative mass function is well-fit by $N(>\msub)\propto 
\msub^{-1}$ down to $\msub=4\times 10^6\,\msun$. Sub-substructure
is apparent in all the larger satellites, and a few dark matter lumps are now resolved
even in the solar vicinity. In Via Lactea, the number of dark satellites
with peak circular velocities above $5\,\kms$ ($10\,\kms$) exceeds 800 (120).  As shown
in Figure \ref{VLsat}, such finding appears to exacerbate the so-called ``missing satellite problem'', the 
large mismatch between the twenty or so dwarf satellite galaxies observed 
around the Milky Way and the predicted large number of CDM subhalos (Moore et al.
1999; Klypin et al. 1999).  Solutions involving feedback mechanisms that make halo 
substructure very inefficient in forming stars offer a possible way out (e.g. Bullock et al. 2000; 
Kravtsov et al. 2004; Moore et al. 2006). Even if most dark matter satellites have no
optically luminous counterparts, the substructure population
may be detectable via flux ratio anomalies in strong gravitational lenses
(Metcalf \& Madau 2001), through its effects on stellar streams (Ibata \etal 2002), 
or possibly via $\gamma$-rays from dark matter annihilation in their cores (e.g. Bergstrom \etal 1999;
Colafrancesco \etal 2006).  We are coming
into a new era of galaxy formation and evolution studies, in which fossil signatures
accessible today within nearby galaxy halos will allow us to probe back to early epochs,
and in which the basic building blocks of galaxies will become recognizable in the
near-field. Understanding galaxy formation is largely about understanding the
survival of substructure and baryon dissipation within the CDM hierarchy. 

\vspace{1cm}
\noindent
{\bf Acknowledgements}

I would like to thank my collaborators, J\"urg Diemand and Michael Kuhlen, for 
innumerable discussions and for their years of effort on the science presented 
here. It is my pleasure to thank the organizers of ``The emission line universe"
for their hospitality and patience in waiting for this proceeding, and all 
the students for making this a very enjoyable winter school. Support for this work 
was provided by NASA grant NNG04GK85G.

{}
\end{document}